\documentclass{emulateapj}

\usepackage[intlimits]{amsmath}
\usepackage{amssymb}

\shorttitle{Variational data assimilation in a solar dynamo model}
\shortauthors{Jouve et al.}

\begin{document}


\title{Assimilating data into an $\alpha\Omega$ dynamo model \\
of the Sun: a variational approach}



\author{Laur\`ene Jouve}
\affil{Universit\'e de Toulouse; UPS-OMP; CNRS; IRAP; 14, avenue Edouard Belin, 31400 Toulouse, France}

\author{Allan Sacha Brun}
\affil{Laboratoire AIM Paris-Saclay, CEA/IRFU Universit\'e Paris-Diderot CNRS/INSU, 91191 Gif-Sur-Yvette, France}

\and

\author{Olivier Talagrand}
\affil{Laboratoire de m\'et\'eorologie dynamique, UMR 8539, Ecole Normale Sup\'erieure, Paris Cedex 05, France}




\begin{abstract}
We have developed a variational data assimilation technique for the Sun using a toy $\alpha\Omega$ dynamo
model.  The purpose of this work is to apply modern data assimilation techniques to solar data using a
physically based model. This work represents the first step toward a complete variational model of solar magnetism.
We derive the adjoint $\alpha\Omega$ dynamo code and use a minimization procedure to invert the spatial dependence of 
key physical ingredients of the model. We find that the variational technique is very powerful and leads to encouraging results
that will be applied to a more realistic model of the solar dynamo.
\end{abstract}

%


\def\brr{\mbox{\huge$\vert$}}
\def\intxy{\int_{x_1}^{x_2}\int_{y_1}^{y_2}}
\def\intxt{\int_{x_1}^{x_2}\int_{t_1}^{t_2}}
\def\intyt{\int_{y_1}^{y_2}\int_{t_1}^{t_2}}
\def\vtx{\, dydt \brr_{x_1}^{x_2}}
\def\vty{\, dxdt \brr_{y_1}^{y_2}}
\def\vtt{\, dxdy \brr_{t_1}^{t_2}}
\def\da{\delta A}
\def\db{\delta B_z}
\def\de{\delta \eta}
\def\dv{\delta v}
\def\dh{\delta \alpha}
\def\dj{\delta \mbox{$\cal{J}$}}
\def\jj{\mbox{$\cal{J}$}}
\def\apj{Astrophysical Journal}
\def\apjl{Astrophysical Journal Letters}
\def\aap{Astronomy and Astrophysics}
\def\jgr{Journal of Geophysical Research}
\def\grl{Geophysical Review Letter}
\def\solphys{Solar Physics}
\newcommand{\intt}{\iiint\limits_D}

%
\section{Introduction}

\subsection{Predicting the solar activity}

At its surface, the Sun exhibits a turbulent and very active
  behavior, with magnetic phenomena as diverse as sunspot emergence, flares, prominences, coronal mass ejections (CMEs).
Quite unexpectedly this magnetic activity is cyclic. The full 22-year cycle is
composed of two consecutive 11-year sunspot cycles (producing the
  so-called butterfly diagram). Coexisting with these large-scale ordered magnetic structures 
are small-scale but intense magnetic fluctuations that emerge over much of the solar surface, with
little regard for the solar cycle \cite[see][]{Stix02}. 
It is currently thought, that, in order to explain this activity and
the large diversity of observed magnetic phenomena, the Sun must
operate two conceptually different dynamos: a large-scale/cyclic
dynamo \cite{Moffatt78,Brun04,Charbonneau05}
and a turbulent small-scale one \cite[e.g.,][]{Cattaneo01,Ossendrijver03}. 

This cyclic activity  has been observed directly since the early
1600's and traced back (indirectly) via $^{10}$Be concentration found in ice core 
for at least 10,000 years \cite{Beer98}. This intense activity is known to have
a direct impact on the Earth's upper atmosphere and on our technological society.
Being able to anticipate and predict the turbulent solar dynamics and magnetic activity is thus crucial
if we wish to prevent damages to our satellites or interferences in our communications.
This has led to the development of space weather studies and forecast. Answering key questions such as which physical processes lead to eruptive phenomena, what is the associated spectrum of solar energetic particles (SEP) and
what leads to geoeffective interplanetary coronal mass
ejections (ICMEs) constitute the main purpose of space weather
\cite{Schwenn06}. 

Solar eruptive phenomena are associated with active regions, i.e complexes of sunspots, that possess intricate magnetic field topology.
There is a direct link between internal magnetism and these surface magnetic phenomena, since active regions are related to the emergence of strong
toroidal structures most likely generated in the deep solar tachocline of intense latitudinal and radial shear at the base of the convection zone \cite{Cline03,Browning06,Brun11}. These toroidal
structures become unstable, subsequently rise through the solar convection zone 
to appear at the surface as active regions \cite{Magara03,Fan03,Archontis05,Jouve09} and are advected by convective motions on the solar surface (Wang \& Sheeley 1991). 
However, the exact link between the solar cycle, CMEs and the geoeffectiveness of solar events is not straightforward to assess \cite{Pevtsov01}.
It is however clear that one important goal of space weather is to characterize the configurations (strength, location, field topology, etc...) that lead to geoeffective events. 
One way to progress in our ability to predict solar activity is to assimilate quality observations in modern numerical models of solar inner and outer magnetism (Schrijver \& Derosa 2003).

Hathaway et al. (1999) summarize most of the methods used to predict the next solar cycle
using historical data. Methods such as regression or curve fitting work well near solar maximum while others such as geomagnetic
precursors perform better near minimum.
It has also been empirically determined that odd numbered cycles
are usually stronger than even numbered ones (possibly indicating a preferred orientation of the inner solar magnetic field) and that 
on average the cycle rises in 4.8 years and falls in 6.2
years, even though strong cycles rise faster to their maximum.
A useful quantity to assess the intensity of a cycle is the yearly averaged
Wolf sunspot number:
$$
R = k (10 g + s)
$$
with  $g$ the number of sunspot groups, $s$ the total number of
individual sunspots in all groups
and $k$ a variable scaling factor (with usually $k < 1$) that accounts for instruments or observation conditions.
Hathaway et al. (1999) suggest that a synthesis of current methods can provide a more accurate and useful
forecast of the evolution of the Wolf number. 
 Cycle 23 was predicted by the \emph{solar cycle 23 panel} to be slightly stronger ($R \simeq 160$) than cycle 22.
However with an observed value of about 120, it turned out to be almost as weak as the even numbered cycle 20 ($R= 105.9$ in 1968). 
Further, in the prediction summary of the \emph {solar cycle 23 panel}, only few of the many predictions 
(even by taking into account their error bars), were actually including the observed value of 120. 

One thus needs to be careful with the standard indicators used up to now. The existence of a panel prediction can be seen as an attempt to use ensemble forecasting \cite{Kalnay03}, similar to what is done in meteorology. The relative success of
these methods, in particular for cycles 21 and 22 (much less so for cycle 23) could be a sign that the set of model equations used in the panel form a good 
ensemble. However most of the techniques considered by Hathaway et al. do not resolve the spatial dependence of the solar activity,
they just focus on global properties such as number of sunspots or the timing of the next maximum. As such, these
techniques are much less sophisticated than the ones used in weather forecasting. We thus need to develop more physically based forecast models
of the solar cycle. 
Historically two types of physical models have been developed in order to understand the solar global dynamo: 2-D mean field models
and 3-D magnetohydrodynamic (MHD) simulations \cite{Ossendrijver03}. However none of these models were used, up to very recently, to predict
the evolution of the solar cycle. 
In order to take into account the spatial dependency of the solar activity, more recent approaches solve numerically the induction 
equation in a meridional plane and impose through a surface term the
observed latitudinal band of activity 
\cite{Dikpati06,Cameron07,Nandy11}.
By assimilating sunspot or meridional flow data, they try to predict
the peak and timing of cycle 24. 

Today, the predictions for the
current solar cycle (recently summarized by the \emph{cycle 24 prediction panel}) differ quite significantly from one model to another (Hathaway 2010). 
Some techniques, such as the ones based on geomagnetic precursors, predict a weak cycle 24 
\cite[$R<100$,][]{Svalgaard05,Duhau03,deJager09}, others based on
dynamo models or meridional flow speed predict a stronger  cycle
\cite[$R > 140$, ][]{Dikpati062, Hathaway04}. It is worth noting that all the predictions
for a weak cycle 24 rely on cycle 23, i.e cycle $n$ is correlated with cycle $n-1$,
whereas those predicting a strong cycle 24 (i.e stronger than cycle 23) favour a correlation with cycle 22, i.e cycle $n$ is well correlated with cycle $n-2$.
The predictions of the cycle 24 panel also differ on the timing of the
next maximum. In 2008, the predictions were that the maximum would occur between 2010 and 2012, depending on how fast 
the next cycle would rise to reach its maximum (fast if strong, slow if weak). It is now clear that the maximum will be reached late in 2013 or in 2014, confirming again the difficulty to
predict the solar cycle. Some recent efforts have been undertaken to improve this situation.
Kitiashvili \& Kosovichev (2008) for instance have used
  assimilation of data in solar dynamo models to predict the solar activity \cite[see also the work of][]{Choudhuri07, Roth09, Rempel09}.

Assimilation of solar data in numerical models has thus already started \cite{Dikpati04, Kitiashvili08, Belanger05,Schrijver03}.
However, intrinsic difficulties in the solar weather forecast are linked to the fact that we do not have yet a complete comprehension of the
solar magnetic dynamo, cycle and surface activity. For every ``piece''
constituting the full puzzle, theoretical developments are still 
underway. This work intends to contribute to this effort.

\subsection{Modern data assimilation techniques in weather forecasting}

In meteorological centers, data assimilation has been operational for many decades already. Various approaches have been developed, becoming
more and more sophisticated. Data assimilation can be defined as ``using all available information, to
determine as accurately as possible the state of the atmospheric (or oceanic) flow'' \cite{Talagrand97}. The purpose of the work presented in this paper is to add the 
words 'solar flow and activity' at the end of the quote.

Modern data assimilation techniques rely on statistical estimation theory, such as least squares methods.
The generalization of such statistical methods to multivariate systems, leads to what is called the optimal interpolation (OI) for 
data assimilation \cite{Lorenc81}. Optimal interpolation consists
  in taking into account (assimilating) the new information that the observational data 
provide in order to advance in time the ``background'' state (also called first guess or prior information) that the weather 
forecasting numerical code has predicted. The increment is obtained by taking the difference, or innovation, 
between the observational data and the observation operator.  The new state or analysis is then the 
result of the assimilation/forecast procedure. More specifically, let ${\bf x^b}$ be the background vector state characterizing 
the current state of the model, $H$ the observational operator and ${\bf y^o}$ the observational data to be assimilated in the model, 
then one can show that the analysis ${\bf x^a}$ is:

\begin{equation}
{\bf x^a}={\bf x^b} + W({\bf y^o}-H({\bf x^b})),
\end{equation}
where ${\bf y^o}=H({\bf x^{real}})+error$ and where $W$ represents the weights determined from the estimated statistical error covariances of the forecast and the observations \cite{Kalnay03}.
This equation is the base of modern data assimilation. The various assimilation methods will differ in the exact definition of W. 

\begin{figure}[!ht]
\center
\includegraphics[width=8cm]{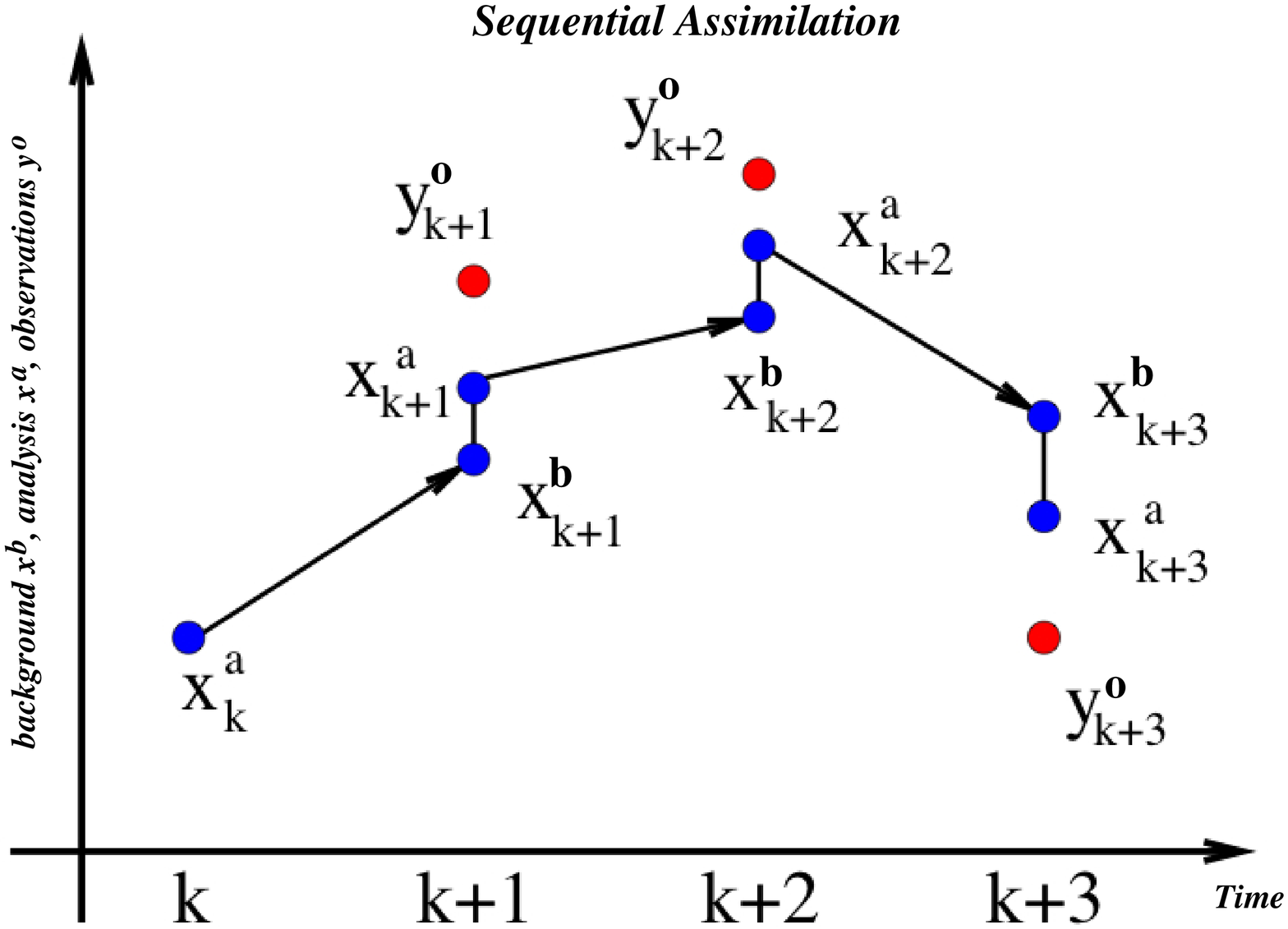}
\includegraphics[width=8cm]{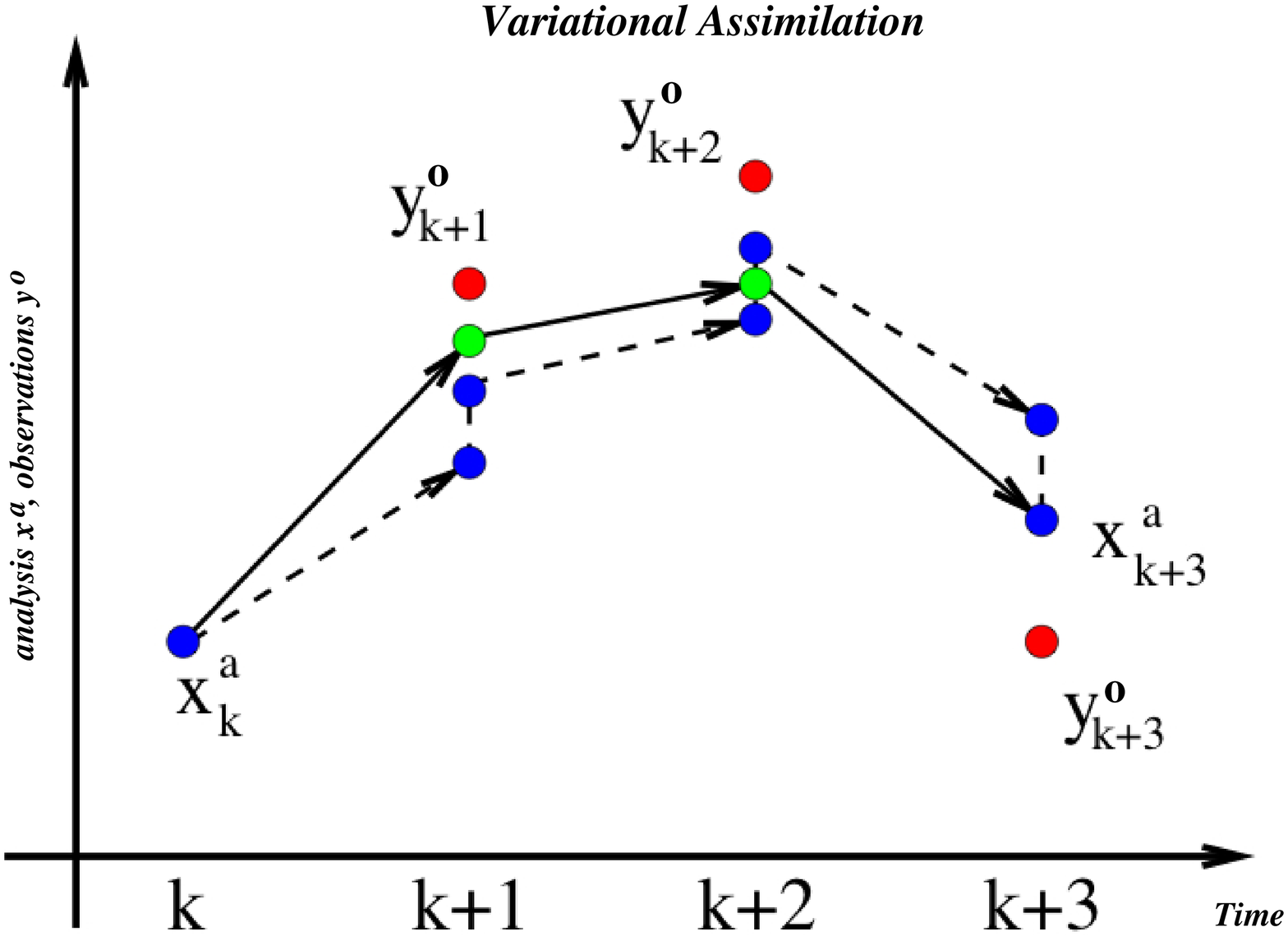}
\caption{Schematic representation of the sequential and 4D variational data assimilation methods used in weather forecast (adapted from Bocquet, 2011). Upper panel: in the sequential method, the background state ($x^b$)
is updated every time observations are available (time between k and k+1) and the model evolves the state until the next step (following the arrows), at which observational data ($y^o$) are again assimilated to produce the analysis ($x^a$). 
Lower panel: 4D variational method and comparison with sequential assimilation. In the 4D variational method, within a time interval the model and the observations are taken into account in the cost function $\jj$ that needs to be ``minimized''. The minimization of this cost function results in a best trajectory (plain arrows) across the observations.}
\label{fig1}
\end{figure}

In practice, the background state, the observations and even the numerical model used to simulate the Earth's atmosphere (i.e 
the primitive equations), possess errors. The assimilation methods
consist in predicting the evolution of the errors and of 
course of minimizing it, i.e keeping it under control as much as possible given the very chaotic nature of the Earth's atmosphere.
Errors in the dynamical atmospheric system are known to double every two to three days, which leads to a 
predictability limit for weather forecasting that Lorenz in 1963 was the first to quantify to be of the order of 15 days. This is a very strong constraint on
our ability to predict weather patterns and solar equivalent predictability limits must exist. However some atmospheric properties may be easier to predict over 
long periods than others, such as weekly averaged rainfall or temperature. It is likely that for the Sun, some
characteristics could also be predicted over 
a longer period of time. 

In order to have a better control of the evolution of the errors,
  data assimilation methods were developed and split into two categories: 
sequential or variational \cite[see][]{Talagrand97, Daley91, Kalnay03}.
As shown in Fig.\ref{fig1}, in the sequential methods, such as OI or Kalman filter, observational data are assimilated in the numerical model at fixed time, say every 6 hours, and then
evolved forward in time. In the so called 4-D variational techniques,
one seeks to minimize a cost (or misfit) function $\jj({\bf \xi})$ (representing the
misfit between the observations and the outputs of the model) within
a certain time interval (usually 12 hours) for which data are available
before making a forecast. The procedure converges when $\jj$ reaches its minimum which occurs for ${\bf \xi}={\bf x^a}$ \cite[see][]{Talagrand03}. 
Then in the next 12-hour periods, the procedure is applied again, using as background state the numerical model of the previous 12 hours. 
The latter technique is the one we wish to apply to the solar
dynamo problem. 

\subsection{Variational assimilation and the adjoint method}

Variational methods require the development and maintenance of a so-called adjoint model 
of the dynamical equations under consideration \cite{LeDimet86}. This adjoint model computes efficiently the gradient $\partial J/\partial \xi$ necessary to the iterative minimizing procedure, 
by evolving backward the adjoint system of equations from the forward temporal integration \cite{Talagrand03, Kalnay03}. 
Such a method is for instance also useful if one seeks to determine the gradient of a variable with respect to a large set of input variables. One can also evaluate the sensitivity of an erroneously predicted feature in the flow in order to assess which input variables are responsible for the error. 

Developing an adjoint model is a straightforward but costly task and no such models have been 
yet developed for the full MHD system of equations (and in particular the induction equation for the magnetic field, see next sections) that is required 
to model the solar dynamics and magnetic activity. The development of the adjoint model of the induction equation is one purpose of this work.

Let us now enter a little bit more into the details of the adjoint
procedure in order to understand how it eases the evaluation of the
gradient of the cost function $\cal J$  with respect to all the input
parameters \cite[see][]{Talagrand91}.

We start by considering a composition of operations  
$G=G_l \odot G_{l-1} \odot ...\odot G_2 \odot G_1$
(where $G$ is a differentiable function) that, given a set of 
input variables $u=(u_1, u_2, u_3, ..., u_{n-1}, u_n)$,
determines a set of output variables $v=(v_1, v_2, v_3, ...., v_{m-1}, v_m)$.

This process can be described by the following equation
\begin{equation}
v=G(u)
\end{equation}
A variation $\delta v$ on the output data leads to a variation $\delta
u$ of the input data that is given at first order by the tangent linear equation:
\begin{equation}
\delta v= G^{\prime}\delta u
\end{equation}
where $G^{\prime}$ if the local Jacobian matrix of G, i.e.

\begin{equation}
G^{\prime}= \left(\frac{\partial v_j}{\partial u_i}\right)_{1 \leq j \leq m, 1 \leq i \leq n}
\end{equation}

Let us now consider a scalar cost function $\cal J$, function of the output
variables $v$. The gradient of the function $\cal J$ with respect to the input variables $u$ reads:
\begin{equation}
\frac{\partial {\cal J}}{\partial u_i}=\sum_{j=1}^{m}\frac{\partial v_j}{\partial u_i} \frac{\partial {\cal J}}{\partial v_j} \,\,\, {\mbox {with}} \,\,\, i=1,...,n
\end{equation}
which is in matrix notation
\begin{equation}
\nabla_{u}{\cal J}=G^{\prime \star} \nabla_{v}{\cal J}
\end{equation}
where $G^{\prime \star}$ corresponds to the transposition of
$G^{\prime}$ (hence the operator represented by the matrix $G^{\prime
  \star}$ is the adjoint operator of the one represented by $G^{\prime}$).

The adjoint method thus allows to compute the gradient of $\cal J$
with respect to the input variables by considering the above
expression (see appendix for more details). Note that since $G$ is the composition of elementary
process  $(G_k)_{k=1,...,l}$, the transpose of the Jacobian
  matrix $G^{\prime}$ will be the product of the transposes of the individual
  Jacobian matrices $G_k^{\prime}$, taken in the reversed order:
\begin{equation}
G^{\prime \star}=G_{1}^{\prime \star} \times G_{2}^{\prime \star} \times ....... \times G_{l}^{\prime \star}
\end{equation}

We have chosen to use this method in the framework of the solar dynamo, by applying it first to a simple  $\alpha\Omega$ mean field dynamo model in Cartesian geometry. 
We give more details in the Appendix on how to apply it
specifically to the induction equation and now describe the model used in this work.


\section{The special case of the $\alpha\Omega$ dynamo}

\subsection{Direct $\alpha\Omega$ dynamo model}

The equation we are interested in is the mean-field induction
equation, derived from the standard induction equation governing the
evolution of a magnetic field in the presence of a conducting fluid
and dissipation, in the framework of mean-field theory. The details of
the derivation of this equation can be found in Steenbeck \& Raedler (1966) or Krause \& Raedler (1980).
The mean-field equation reads: 

 \begin{equation}
\frac{\partial {\bf B}}{\partial t}=\nabla\times({\bf v \times B})+\nabla \times (\alpha{\bf B})-\nabla\times(\eta\nabla{\bf \times B})
\label{eq_mf}
\end{equation}

\noindent where {\bf B} and {\bf v} are respectively the mean magnetic and
velocity fields, $\alpha$ parametrizes the physical process responsible for the regeneration of poloidal field and $\eta$ is the effective magnetic diffusivity.

We choose to work in Cartesian geometry with coordinates $(x,y,z)$,
which would respectively correspond in spherical geometry to the radius, latitude and longitude. The 3 components of the magnetic field depend only on the $x$ and $y$ coordinates. The domain is defined as $[x_1,x_2] \times [y_1,y_2]$, with a
regular grid spacing assuming $N_x=N_y=30$. For simplicity we assume that $x_{1,2}=\pm1$ and $y_{1,2}=\pm1$. We note that the discretization in $y$ is symmetric with respect to the equator defined by $y=0$.
The poloidal/toroidal decomposition of the magnetic field then reads:

\begin{equation}
{\bf B}(x,y,t)= \nabla\times(A(x,y,t) {\bf e_z}) + B_z(x,y,t) {\bf e_z}
\end{equation}

Reinjecting this poloidal/toroidal
decomposition in our mean-field induction equation, we get two coupled
partial differential equations, one for the poloidal potential
$A$ and the other for the toroidal field $B_z$.

\begin{equation}
\frac{\partial A}{\partial t}=\alpha B_z + \eta (\frac{\partial^2 A}{\partial x^2}+ \frac{\partial^2 A}{\partial y^2})
\label{eq_A_simple}
\end{equation}

\begin{equation}
\frac{\partial B_z}{\partial t}=\frac{\partial v}{\partial x}\frac{\partial A}{\partial y}-\frac{\partial v}{\partial y}\frac{\partial A}{\partial x}+\eta (\frac{\partial^2 B_z}{\partial x^2}+ \frac{\partial^2 B_z}{\partial y^2})
\label{eq_B_simple}
\end{equation}

We choose to neglect the $\alpha$-effect in the equation for the toroidal field since the shear is considered to be the dominating source term. We thus consider a simple $\alpha\Omega$ dynamo model here.
For boundary conditions, we assume for simplicity that both $A$ and $B_z$ are set to zero on the borders $x=x_1$ or $x_2$ for all $y$ and on the borders $y=y_1$ or $y_2$ for all $x$ at all times t. As initial conditions, we choose a dipolar field structure, $A$ being symmetric with respect to the equator $y=0$ and $B_z$ being zero everywhere.

The prescribed velocity field simply expresses as

\begin{equation}
{\bf v}= \Omega_0 x \sin(\pi \frac{y+1}{2}) {\bf e_z}
\end{equation}

\noindent where $\Omega_0$ represents the rotation rate of our domain.

We now need to give the expression for the $\alpha$-effect,
responsible for the regeneration of poloidal field. We choose it to be
antisymmetric with respect to the equator, as is assumed in the Sun from surface kinetic helicity measurements (Komm et al. 2007, 2008) and 3D simulations of the convective interior \cite{Miesch00, Brun04}. Its expression is the following

\begin{equation} 
\alpha=\alpha_0 \cos (\pi \frac{y+1}{2}) 
\end{equation}

\begin{figure}[h]
  \centering
  \includegraphics[width=8cm]{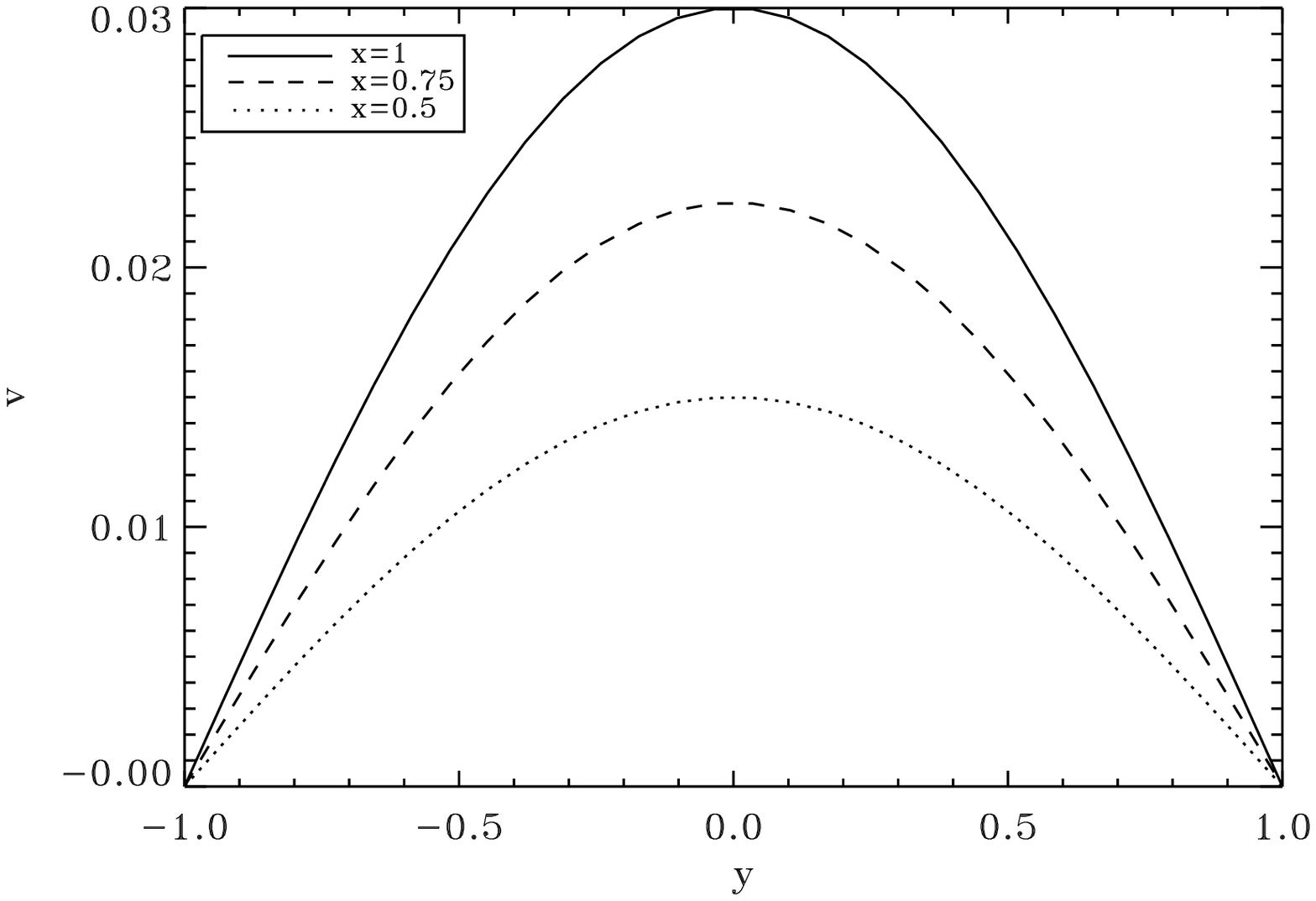}
  \includegraphics[width=8cm]{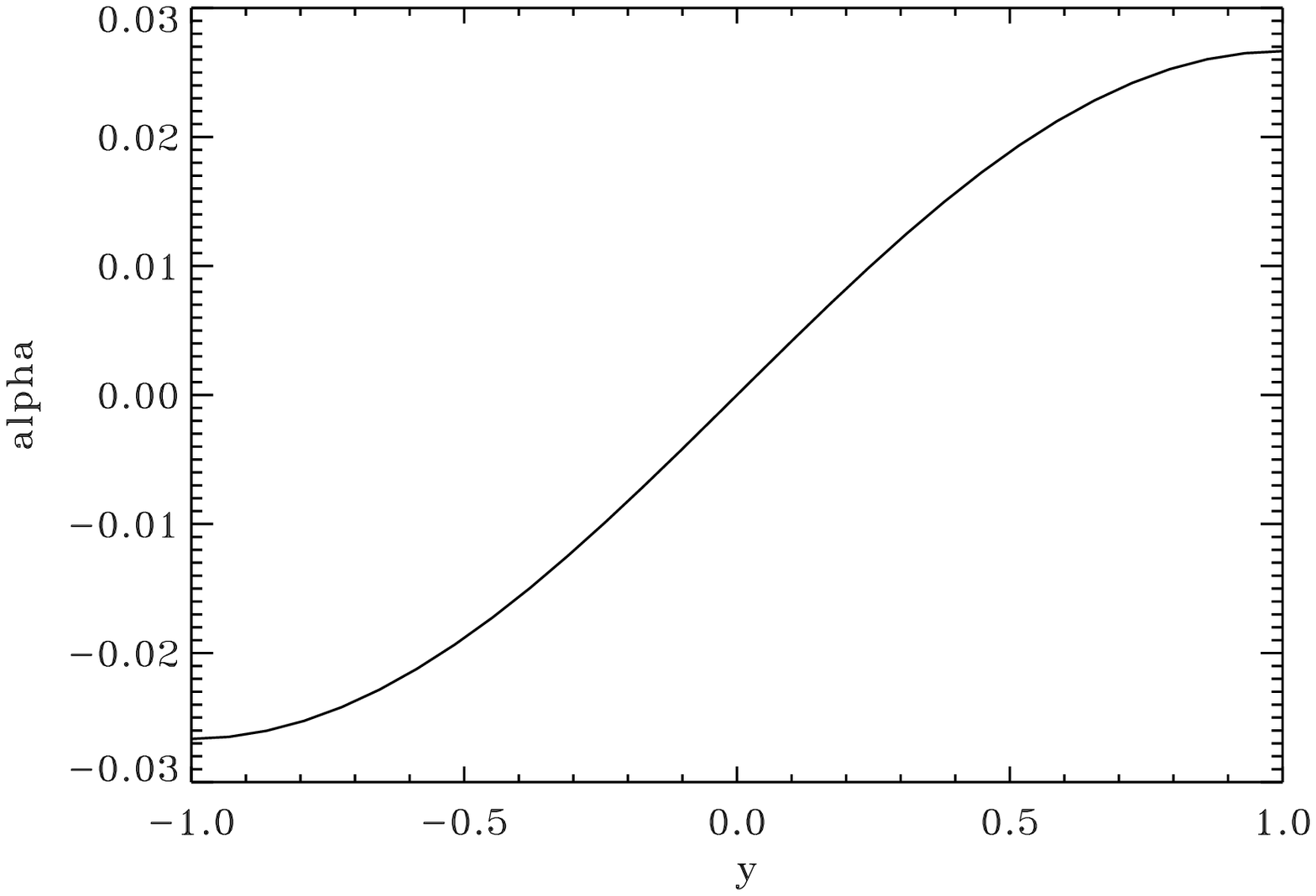}
  \caption{Profiles of $v$ (upper panel) and $\alpha$ (lower panel) used in this
    simple model.}
  \label{fig_ing}
\end{figure}

Finally, the magnetic diffusivity is assumed to be constant
$\eta=cst$. The profile of the physical ingredients of the model are shown in Fig. \ref{fig_ing}.

We can now nondimensionalize those equations by choosing a length scale $L$
and a temporal scale  $L^2/\eta$. This procedure leads to the
definition of physically relevant dimensionless parameters and to the new equations:

\begin{equation}
\frac{\partial A}{\partial t}=C_\alpha B_z + (\frac{\partial^2 A}{\partial x^2}+ \frac{\partial^2 A}{\partial y^2})
\label{eq_A_ad}
\end{equation}

\begin{equation}
\frac{\partial B_z}{\partial t}=C_\Omega (\frac{\partial v}{\partial x}\frac{\partial A}{\partial y}-\frac{\partial v}{\partial y}\frac{\partial A}{\partial x})+ (\frac{\partial^2 B_z}{\partial x^2}+ \frac{\partial^2 B_z}{\partial y^2})
\label{eq_B_ad}
\end{equation}

\noindent with $C_\alpha=\alpha_0 L/\eta$ and $C_\Omega=\Omega_0
L^2/\eta$ the Reynolds numbers measuring the intensity of the $\alpha$
and $\Omega$ effects compared to the Ohmic dissipation. The product of
those two numbers will have to be above a given threshold for
dynamo action to occur.

\subsection{The numerical method and choice of model parameters}
\label{sect_model}

Equations \ref{eq_A_ad} and \ref{eq_B_ad} are solved numerically using
a finite difference scheme in space and time. More specifically, we use a first
order explicit Euler scheme for time integration and a 2nd order centered scheme in space. We thus have to carefully check the CFL condition: the time step will be constrained by the minimum of the advective timescales (related to $\alpha_0$ and $\Omega_0$) and the diffusive time (related to $\eta$).
The output of such simulations will be two 3D arrays $A$ and $B_z$ (two dimensions in space and one in time) of dimension $30\times 30\times 1000$.

\begin{figure}[h]
  \centering
  \includegraphics[width=8cm]{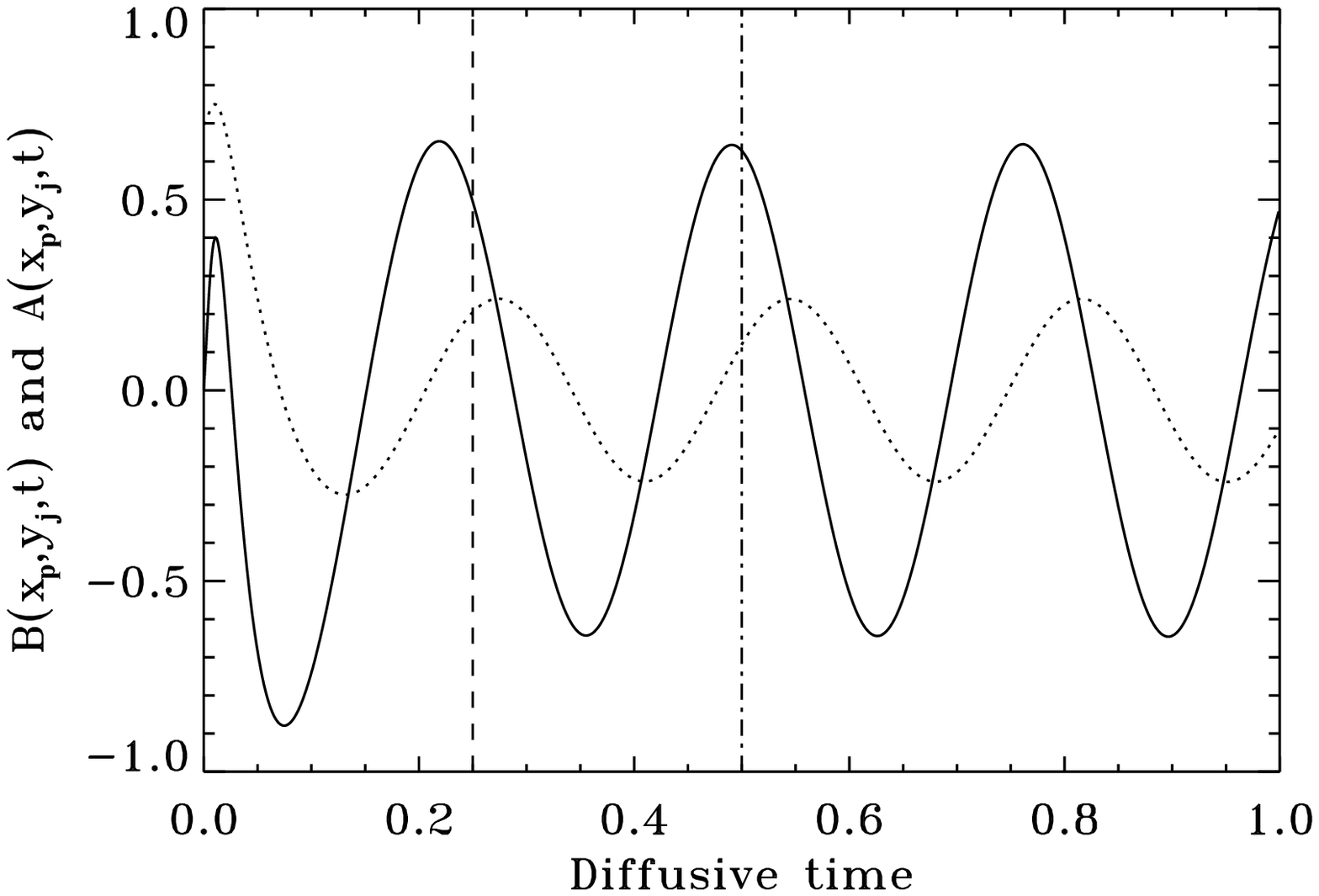}
  \includegraphics[width=8cm]{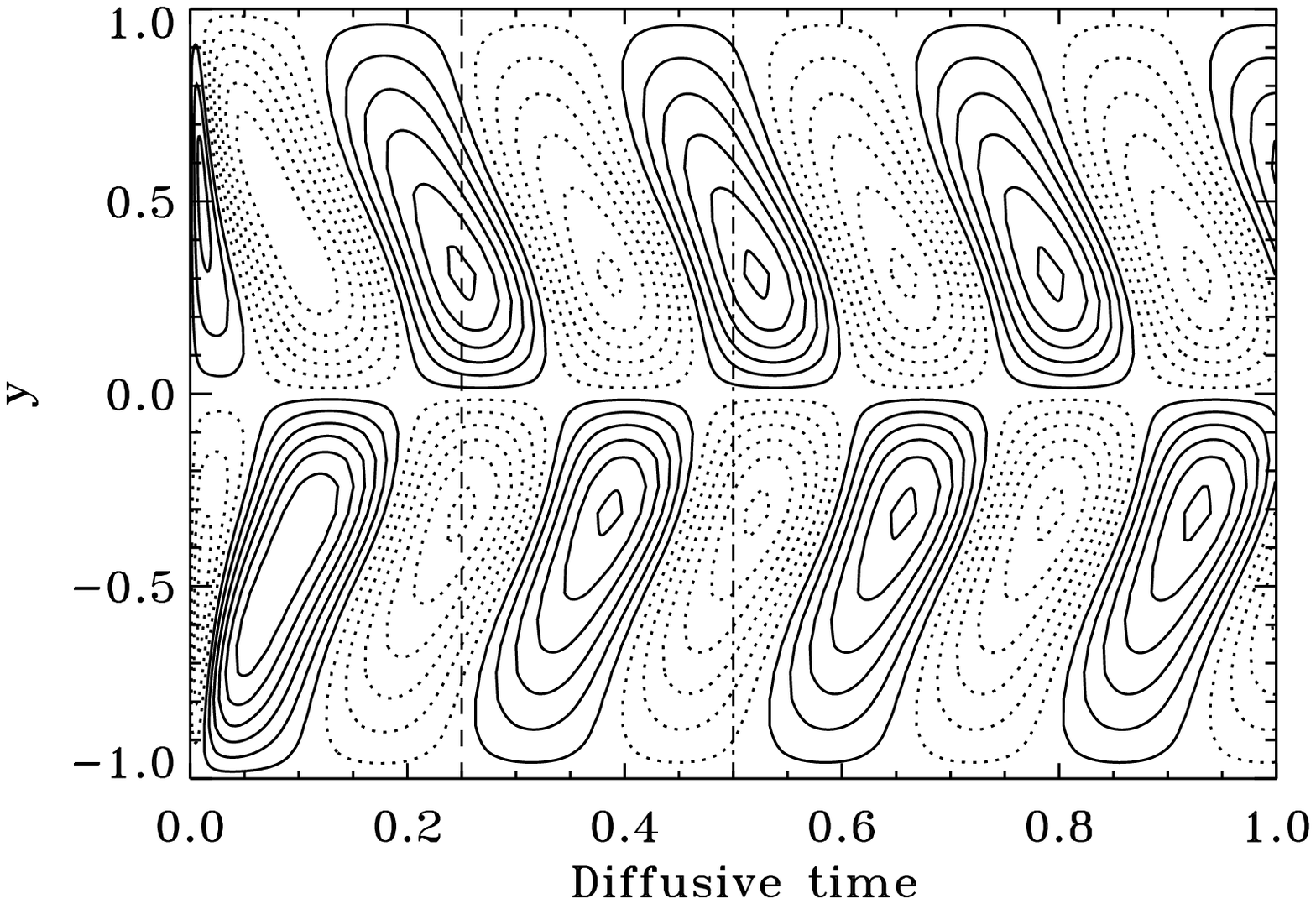}
  \caption{Representative case for $\alpha_0=-0.02665$, $\Omega_0=0.03$
    and $\eta=0.001$: time evolution of the toroidal field (plain line) and poloidal potential dotted line at a
    particular point in space (upper panel) and time-latitude cut of the toroidal
    field at a depth $x_p$ near the top of the domain (lower panel). The latter represents the
    butterfly diagram of our solution. The dashed and dashed-dotted lines represent respectively the end of the first and second assimilation window.}
  \label{fig_solution}
\end{figure}

A typical dynamo solution found in our model is shown in Fig.
\ref{fig_solution}. Our set of parameters ($\alpha_0=-0.02665$, $\Omega_0=0.03$
    and $\eta=0.001$) was carefully chosen so that we are in the
    marginally stable regime. We are exactly at the threshold for
    which the dynamo instability is triggered, i.e. the growth rate of
    the instability is purely imaginary and the fields oscillate
    around zero without growing. If the absolute value of the dynamo numbers were further
    increased, the dynamo instability would grow and in this linear
    case, the magnetic energy would increase exponentially without bound.

The lower panel of Fig.\ref{fig_solution} shows the butterfly diagram, i.e. a time-latitude cut
of the toroidal field $B_z$ at a particular location in depth. Again, our
    choice of parameters, especially the sign of the dynamo number
    $C_\alpha C_\Omega$ was made to produce an equatorward
    propagating dynamo wave. Indeed, Yoshimura (1975) showed that the direction of propagation of
    the dynamo wave when a radial shear is present depends on the sign
    of the product $\alpha_0\Omega_0$.

\subsection{Generating observational data}
\label{sect_twin}

The idea of this work is to show that data assimilation techniques can
be applied to solar dynamo models. To do so, we develop the adjoint
model necessary for the variational assimilation described in Section
1 and we test its  validity. We will generate
synthetic observations with a certain set of parameters and will use
our adjoint model to minimize the cost function
and recover the right parameters starting from a random initial guess.
Such a procedure is called a twin experiment and has been used in
various situations and studies before \cite[e.g.][]{Fournier07}.

We choose as our synthetic data the dynamo solution presented in the
previous section. In our twin experiments, the observations are
chosen to be the toroidal field $B_z$ at $ny$ specific points in space and $nt$ points in time, corresponding in the Sun to the value of the sunspots magnetic
field at different latitudes and time during the cycle.

 The aim of the adjoint procedure will then be to reconstruct the state vector $\hat{\alpha}$, the dimension of which is the number of points in the $y$-direction, fixed to $30$ in all calculations.
In the remaining of the paper, we distinguish the true physical ingredient (denoted $\alpha$) and the state vector to be reconstructed (denoted $\hat{\alpha}$).

\subsection{Adjoint $\alpha\Omega$ dynamo model}

In the appendix, we present the derivation of the continuous adjoint induction equation. This helps us
gaining some insight on the relation between the mathematical definition
of an adjoint operator and the procedure we are using
in this work. However, it has to be pointed out that it is not the
adjoint partial differential equation which will be discretized to build the adjoint code.
To do so, we rather attribute an adjoint
instruction to each direct instruction in the tangent linear model deduced from the linearization of the direct model. This follows the formal procedure described in \cite{Talagrand91} and \cite{Giering98}.

The goal of the whole variational experiment here is to minimize a cost
function $\jj$ which will measure the misfit between the observations and
the values of the variables calculated by the numerical model. To do
so, we need first to define a proper cost function which
will have to be minimized. Secondly, the idea is to choose a minimization
algorithm which uses the values of the cost function (calculated
by the direct integration of the model) and its gradient with respect
to all input parameters (produced by the adjoint integration).

For our studies, we choose the following cost function

\begin{equation}
\jj=\sum_{k=1}^{nt} \, \, \sum_{j=1}^{ny} \, \, \frac{(B_z(x_p,y_j,t_k)-B_z^{obs}
(x_p,y_j,t_k) )^2}{\omega(j,k)^2}
\end{equation}

\noindent where $x_p$ is a particular depth. It is chosen to be
close to the boundary of the domain in our case, in the attempt to get
closer to the real Sun where data are available only at the
surface. $\omega(j,k)$ can be adjusted to give more or less weights to
some observations, if for example some are more reliable than
others. This would happen if a new instrument with more accuracy was
launched (then we can expect the errors on the observations to vary in
time) or if observations of certain regions in space were less subject
to uncertainty. In our twin experiments described below, $\omega(j,k)$ is chosen to be constant,
i.e. independent on the position in space or time.

The cost function is then minimized through a quasi-Newton method
which uses the first and second derivatives of the function.
 A particularity of the quasi-Newton methods is that they
need the gradient of the function (which is here provided by the
adjoint integration) but do not require exact computation of the Hessian
matrix, which is instead approximated by an iterative algorithm (here
the formula of Broyden-Fletcher-Goldfarb-Shanno is used to update the
value of the Hessian approximation). See Polak (1971) for details
about the algorithm. 
We note here that the computation of the gradient of the cost function
through the adjoint code has been tested. To do so, we checked that the quantity
\begin{equation}
\jj(X+\delta X)-\jj(X)-\delta X \cdot \nabla \jj(X)
\end{equation}
with $\nabla \jj(X)$ calculated through the adjoint code, is order $o(\delta X)$ to computer accuracy.

\section{Twin experiments and results}

As discussed in Sect.\ref{sect_twin}, we wish to reconstruct the true state $\alpha$. To do so, we perform several experiments to assess the sensitivity and quality of the reconstructed state.

\subsection{Regular sampling in space}

Our first experiment consists in producing data with the choice of parameters quoted above at regularly-spaced locations in
space and time. More specifically, we fix the value for the $x$-coordinate
(representing the depth in the convection zone) and we produce
observations both in the Northern and Southern hemispheres, with a
regular spacing. Moreover, those observations will be available during
the first cycle(s) only, with a regular spacing in time.

The initial guess is $\hat{\alpha}=0$ on every grid points except at the
boundaries $y=-1$ and $y=1$ where $\hat{\alpha}$ is set to the true state. Indeed, the values of $\hat{\alpha}$ at the boundaries will not affect our cost function since the magnetic field $B_z$ is exactly set to zero at those points (see lower panel of Fig.\ref{fig_solution} and Eqs.\ref{eq_A_simple} and \ref{eq_B_simple}). As a
consequence, in the minimization procedure, only $\hat{\alpha}$ within the
domain is adjusted to reduce the amplitude of the cost function.  
The tolerance on the gradient is set to $10^{-12}$, which is
typically reached after about 300 iterations of our minimization
algorithm. By that time, depending on the number of observations used,
the final value of the cost function varies between $10^{-17}$ and
$10^{-27}$, i.e. has decreased by at least 16 orders of magnitude. 
We note here that the number of iterations might seem large compared to the dimension of the state vector. However, close to both the boundaries and the equator, the amplitude of the toroidal field $B_z$ is about 100 times smaller than at mid-latitude. Since the $\hat{\alpha}$ function only affects the cost function through its product with $B_z$ (see Eqs. \ref{eq_A_simple} and \ref{eq_B_simple}), the recovery of $\hat{\alpha}$ will be less efficient in the regions where $B_z$ is close to zero. If, on the contrary, those points are removed from the assimilation procedure and initially set to their true values, the convergence is much faster (not shown). We will discuss the difficulties of recovering $\hat{\alpha}$ in the equatorial regions in the following sections. 

 \begin{figure}[h!] 
  \centering
 \includegraphics[width=8cm]{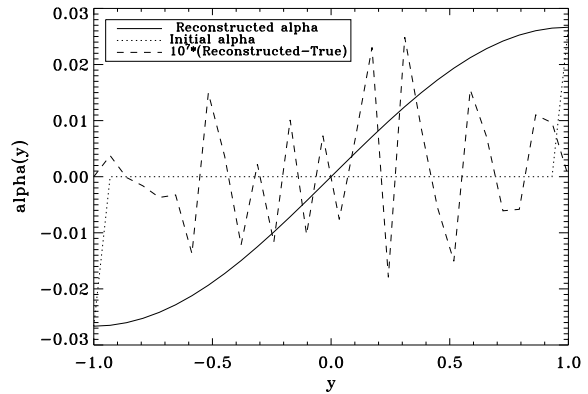}
 \caption{Initial guess and $\hat{\alpha}$ recovered by the minimization of the cost function with
   10 observations in time and 10 regularly spaced observations along
   the y-direction for each of those 10 points in time. We also show
   the error between the recovered $\hat{\alpha}$ and the true state, magnified
   by a factor $10^7$.}
 \label{fig_alpha_reg}
 \end{figure}

We run our minimization
procedure and compare the results obtained when various numbers of observations are
assimilated. The number of points in time can be 5 or 10, located in
the first or first two cycles (see the 2 assimilation windows in Fig.\ref{fig_solution}). In space
(more specifically in the direction of $y$, representing the
latitude), the number of observations varies from 6 to 14. The total
number of observations thus extends from 30 to 140 depending on the
calculation. Figure \ref{fig_alpha_reg} shows a typical result of the
minimization algorithm for 100 assimilated observations. The function
is perfectly recovered and the pointwise error has been reduced by a
factor $10^7$ compared to the initial guess.

\begin{figure} 
 \centering
 \includegraphics[width=8.5cm]{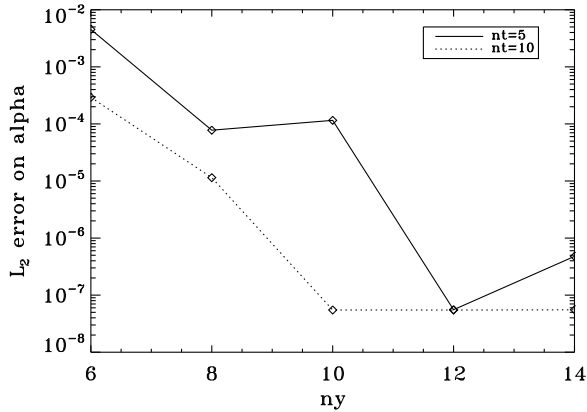}
 \caption{$L_2$ error on $\hat{\alpha}$ (compared to the $\alpha$ used to produce
   the observations) for various numbers of observations in space and
   time. Note the monotonous decrease in the error when 10 points in time are used
   as observations.}
 \label{fig_l2}
 \end{figure}

The first conclusion which can be drawn from this experiment is that, as must necessarily be,
increasing the number of observations decreases the error made on the
reconstructed $\hat{\alpha}$ (see Fig.\ref{fig_l2}). However, even 30 observations in total (5 in
time times 6 in space) are sufficient to get an $\hat{\alpha}$ function
indistinguishable from the true state. The only quantitative way to
compare the different experiments is thus to look at the $L_2$ errors
 between the $\hat{\alpha}$ coming from the minimization
algorithm and the true $\alpha$ used to produced the observations. More
precisely, we calculate the following quantity

\begin{equation}
e=\sqrt{\frac{\sum_{j=1}^{ny} \, \, \left (\hat{\alpha}(y_j)-\alpha(y_j)
    \right)^2}{\sum_{j=1}^{ny} \, \, \alpha(y_j)^2}}
\end{equation}

\begin{figure}[h!] 
 \centering
 \includegraphics[width=8.5cm]{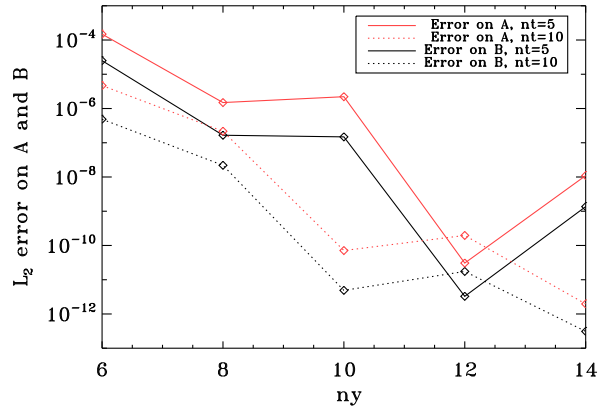}
 \caption{$L_2$ errors on the toroidal (black lines) and poloidal
   fields (red lines) for the different experiments.}
 \label{fig_error}
 \end{figure}

The amplitude of those errors are shown in Fig. \ref{fig_l2}, as a
function of the number of observations in the $y$-direction. For
completeness, we show the results obtained when observations
are located both in the first cycle ($nt=5$) and in the first two cycles ($nt=10$). We clearly show here that the
error almost monotonically drops
when more and more observations are assimilated, reaching values of the order of $10^{-7}$ for the best cases. The reconstructed
$\hat{\alpha}$ then produces poloidal and toroidal magnetic fields very much in agreement with
our synthetic observations, as shown in Fig. \ref{fig_error}.

Figure \ref{fig_error} shows the $L_2$ errors, on the toroidal and
poloidal fields produced by the reconstructed $\hat{\alpha}$ effect, for
various numbers of assimilated observations. Again,
we find a very good agreement both for the poloidal and toroidal
fields even for the smallest number of observations. For larger numbers of observations, the relative errors reach values close to
$10^{-10}$ and even $10^{-12}$ for the toroidal field. It is interesting
to note that the errors on the toroidal field are systematically about one order of magnitude
less than the errors on the poloidal field. This is likely due to the fact
that observations are available on the toroidal field only (e.g. the cost function depends exclusively on $Bz$) and thus a
better agreement is to be expected. We can also note on this figure
that the errors do not grow in time and thus that the functions are
recovered on the whole time interval, even if observations were only
available on the first cycles. This feature is mainly due to
the fact our system of equations is stable to perturbations of the initial conditions, meaning that an initial perturbation would not be amplified nor damped.

 \begin{figure}[h!] 
 \centering
 \includegraphics[width=8cm]{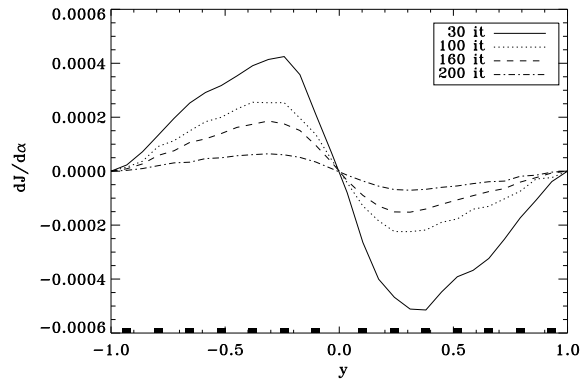}
 \caption{Gradient of $\jj$ with respect to $\hat{\alpha}$ in the case where observations (represented by the squares at the bottom of the graph) are regularly spaced in $y$. The various curves represent the value of the gradient after 30 iterations of the minimization algorithm, $50\times\nabla\jj$ after 100 iterations, $500\times\nabla\jj$ after 160 iterations and $10000\times\nabla\jj$ after 200 iterations.}
 \label{fig_djdar}
 \end{figure}

For the best case considered ($nt=10$, $ny=14$), we found it instructive to follow the evolution of the gradient of the cost function with respect to $\hat{\alpha}$ during the minimization procedure. We choose particular steps in the iterative minimization procedure, separated by sufficiently large decreases of the norm of the gradient. The results are shown in Fig.\ref{fig_djdar} where the gradient is plotted at those steps, with respect to the y-coordinate. The first thing we note is the clear decrease in the amplitude from the beginning to the end of the procedure, the last step chosen (after 200 iterations) being very close to the total number of iterations required to achieve convergence (211 in this case). The second striking property of the curves shown on this figure is the shape of the function, antisymmetric with respect to the equator. This characteristic indicates that the cost function $\jj$ is not sensitive to the values of $\hat{\alpha}$ close to the equator and explains why the difficulties to reproduce the true $\alpha$-effect lie mostly in the equatorial regions. This will be even more obvious in the following sections where data are chosen not to be distributed over the whole domain or when data are perturbed by a random noise. However, the profile of the gradient is not surprising if we consider Eq. \ref{eq_dj} of Appendix \ref{ap_B} and Eqs. \ref{eq_A_simple} and \ref{eq_B_simple}, that clearly demonstrate that if $B_z$ is zero,  $\hat{\alpha}$ has no influence in the equation for the magnetic field. Stated otherwise, the profile of $\nabla_\alpha\jj$ follows that of the mean value of $B_z$ over the time interval in which the assimilation procedure is applied. As a test, we plotted $<B_z(x,y)>_{t}$ with respect to $y$ at a particular point in $x$ (not shown) and indeed, we recovered the exact same profile as what is shown in Fig.\ref{fig_djdar} for the various curves.

\subsection{Irregular sampling in space}

We chose here as observations a quantity $B_z$ related to the intensity of the sunspots magnetic field. In the real Sun, sunspots emerge at mid-latitudes at the beginning of the magnetic cycle and closer and closer to the equator as the cycle proceeds. For a more realistic experimental setting, we have studied different cases for which we have assimilated observations in restricted latitudinal bands. We first show the results of an experiment where data
were available in one hemisphere only and in the next section, we
investigate the case where observations are assimilated in the
activity belt only, i.e. at low latitudes in both hemispheres.

\subsubsection{One hemisphere only sampling}

In this first case, we produce synthetic data only in the Southern
hemisphere (for negative values of $y$) and study the reconstruction
of the $\hat{\alpha}$ function through the minimization algorithm. Again, the
initial guess is $0$ everywhere except on the boundaries and
observations are equally spaced in time and on the first two
cycles only (10 points in time are used here).

 \begin{figure}[h!] 
 \centering
 \includegraphics[width=8cm]{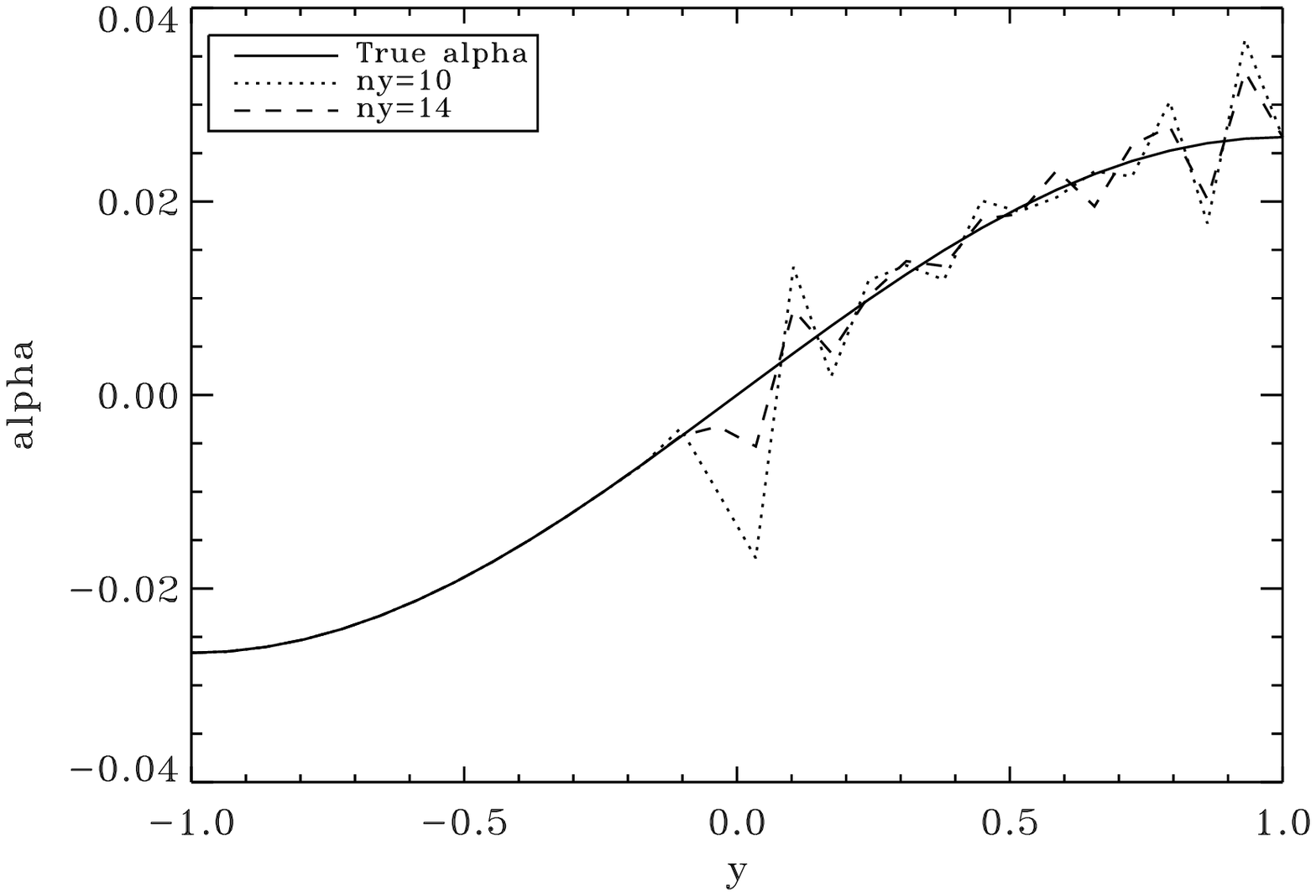}
\includegraphics[width=8cm]{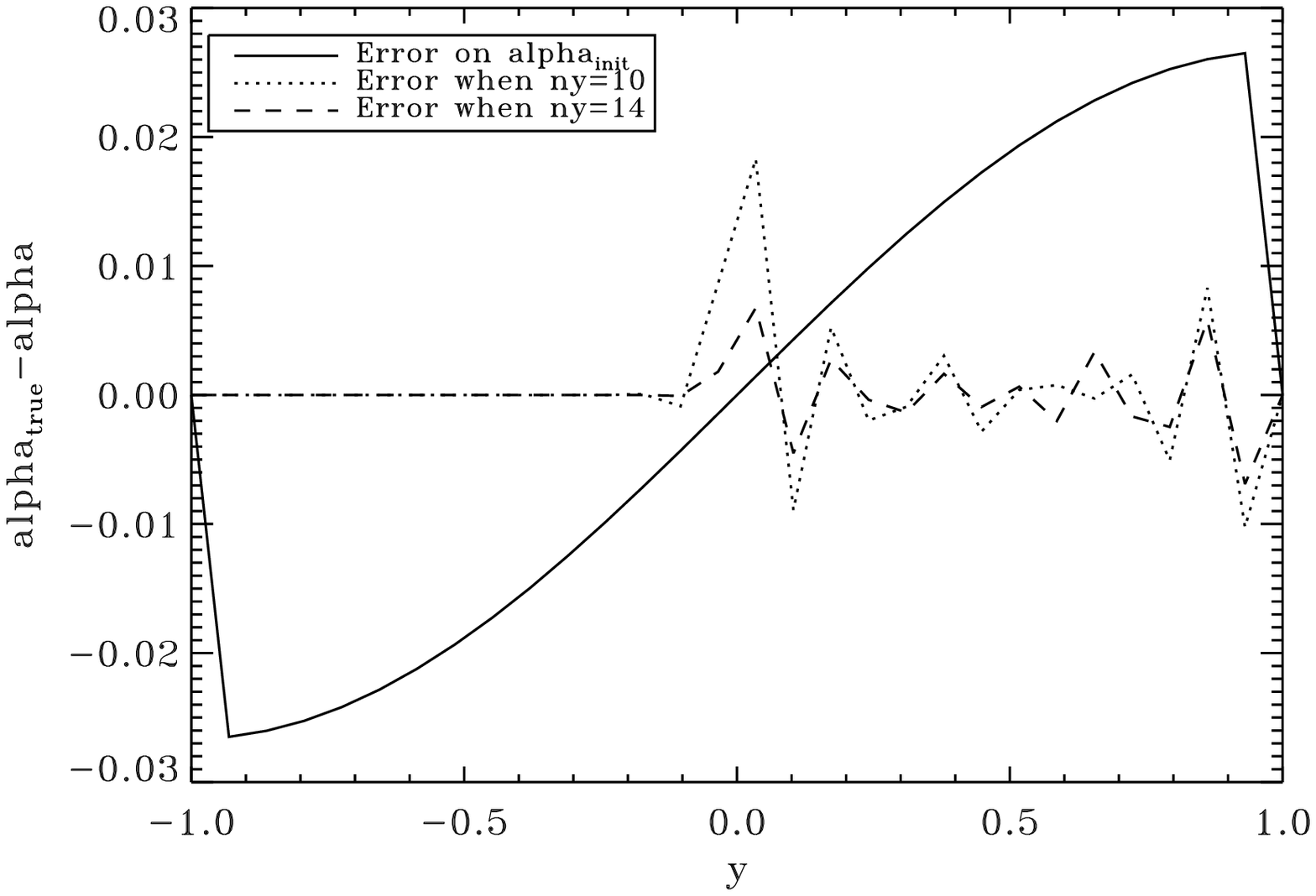}
\caption{ Upper panel: $\hat{\alpha}$ reconstructed after assimilation of data in the
  Southern hemisphere only, with 2 different sets of observations, superimposed with the true state.
Lower panel: Errors made on the reconstructed $\hat{\alpha}$ for the initial guess (see Fig. \ref{fig_ing}) and after assimilation of the 2 sets of observations.}
 \label{fig_alpha_irreg}
 \end{figure}

Figure \ref{fig_alpha_irreg} shows the results of the minimization. It is clear that where data have been assimilated, the reconstruction of the function is
much more accurate than on the Northern hemisphere where observations were
absent. The behavior of the function is much smoother
in the Southern hemisphere and very similar for both sets of
observations. On the contrary, the function strongly fluctuates on the
data-free region and especially in the equatorial region for the
experiment where only 10 points in space were used. However, when observations are added mostly close to the equatorial region, the error is reduced even on the data-free region and the equatorial region is almost correctly recovered.

 \begin{figure}[h!] 
 \centering
 \includegraphics[width=8cm]{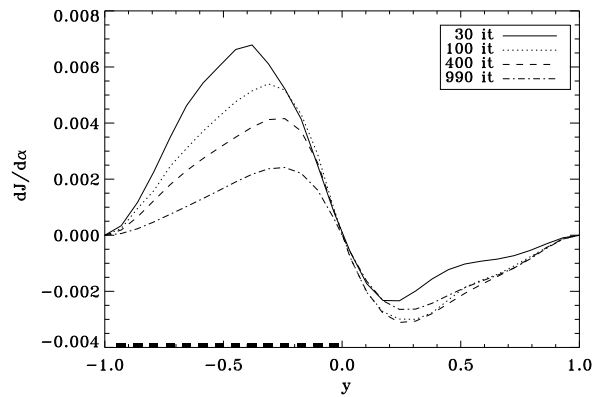}
 \caption{Gradient of $\jj$ with respect to $\hat{\alpha}$ in the case where observations (represented by the filled squares at the bottom of the graph) are available only in one hemisphere. The various curves represent the value of the gradient after 30 iterations of the minimization algorithm, $80\times\nabla\jj$ after 100 iterations, $2000\times\nabla\jj$ after 400 iterations and $20000\times\nabla\jj$ after 990 iterations.}
 \label{fig_djdai}
 \end{figure}

However, even if there exists a clear asymmetry between the two
hemispheres here, it has to be noted that the error on the $\hat{\alpha}$
function after minimization is much less than the initial error, even
in the data-free region. This is shown on the lower
panel of Fig. \ref{fig_alpha_irreg}, where the pointwise error is plotted
for the initial guess and for the recovered $\hat{\alpha}$. We thus conclude
from those experiments that a knowledge of the toroidal field only in
one hemisphere also gives us some information on the profile of the
$\alpha$-effect in the other hemisphere.
 This result shows that a link exists between the two hemispheres, due to various physical processes, explaining why the intensity of the magnetic field in one hemisphere will influence the other hemisphere. In the Sun, this link could be related to magnetic flux crossing the equator at particular moments during the cycle or to the dipolar topology of the poloidal field.

Again, as in the previous section, we have followed the evolution of the gradient of the cost function with respect to $\hat{\alpha}$. Various instants in the minimization algorithm were chosen, namely after 30, 100, 400 and 990 iterations (the larger number of iterations being due to the slower convergence of the algorithm). At the beginning of the minimization procedure, an asymmetry between the two hemispheres is clearly visible, as can be expected.  This is seen in the analysis of the full curve of Fig.\ref{fig_djdai}, which represents the gradient after 30 iterations of the algorithm. The peak value of the gradient in the Southern hemisphere is here about 3 times higher than the peak value in the Northern  hemisphere. However, as the minimization proceeds, the gradient in the Southern hemisphere is reduced more than in the Northern hemisphere, leading to a more and more symmetric profile with respect to the equator.

\begin{figure}[h!] 
 \centering
 \includegraphics[width=8cm]{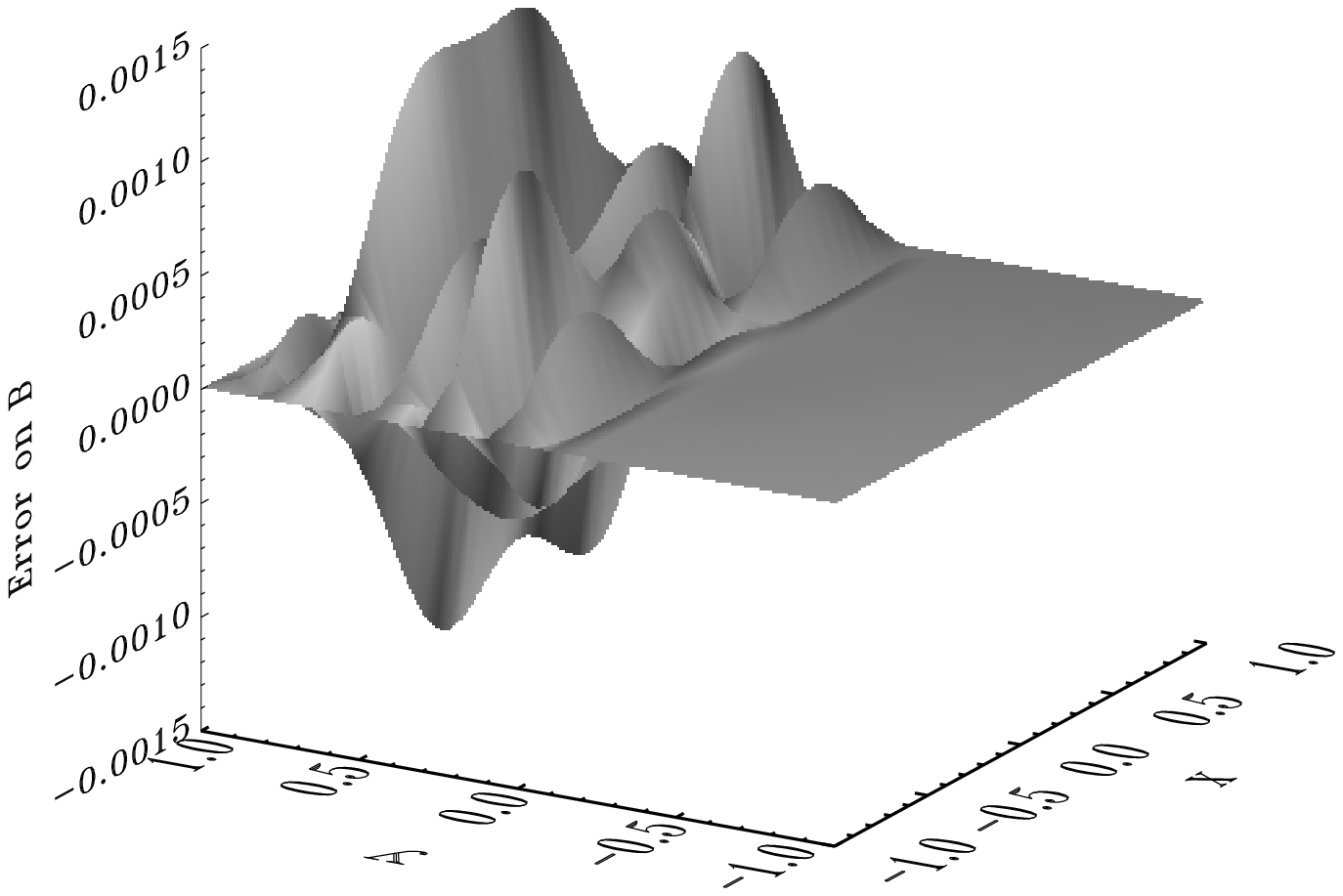}
 \includegraphics[width=8cm]{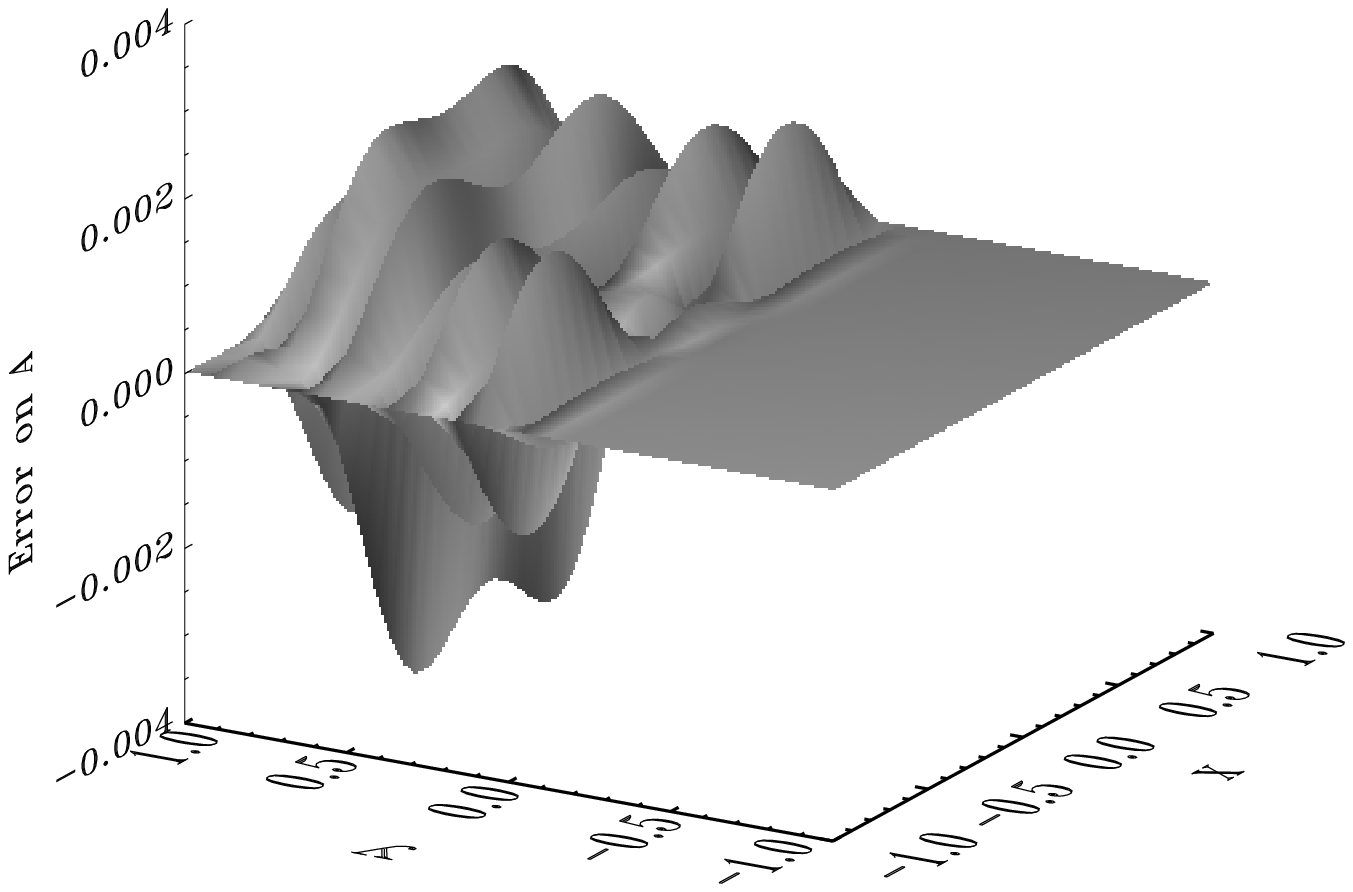}
 \caption{Difference between the components of the toroidal magnetic field (upper panel) and poloidal potential (lower panel) produced by the
   reconstructed $\hat{\alpha}$ and the true state at t=0.5 (in the middle of the time interval).}
 \label{fig_irreg}
 \end{figure}

Once again, we can analyze the quality of the magnetic fields produced
by the reconstructed $\hat{\alpha}$ and calculate its errors
compared to the true state. This is shown in Fig.
\ref{fig_irreg} at one instant, for the case where 8
observations were used. We wish here to focus on the errors at a particular
instant in the simulation since we are interested
in the spatial distribution of the error, rather than on its time
evolution. We show on this figure that again the field
is in very good agreement with the true state in the region where
observations were assimilated, the relative errors reaching values as
low as $10^{-6}$ in these regions. On the contrary, the agreement in
the Northern hemisphere is much worse, even if the relative error is
of the order of $10^{-3}$ for the toroidal field. For
the poloidal field, the errors are again almost one order of magnitude
higher, still due to the fact that observations are available on the
toroidal field only. We should note that the agreement for the
poloidal field 
on the Southern hemisphere is very satisfactory, stressing the
efficiency of the variational assimilation. 

\subsubsection{Active latitude band sampling}

If we choose as observations the sunspot magnetic field detected
during solar cycles, we have to be aware that observations
will mainly be available in the solar activity belt, i.e. between
about $-35^o$ and $35^o$ in latitude (Hathaway 2010). 

 \begin{figure}[h!] 
 \centering
 \includegraphics[width=8cm]{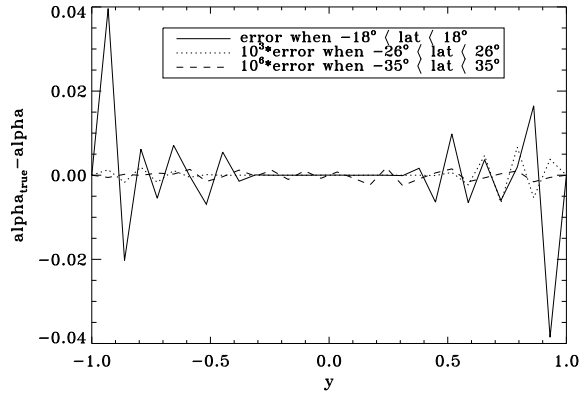}
\caption{Errors made on the reconstructed $\hat{\alpha}$ after assimilation of various
  numbers of observations located in the equatorial regions, between
  $-35^o$ and $35^o$ for the broadest interval.}
\label{fig_alpha_act1}
\end{figure}

We thus choose to investigate
the recovery of our true state in a case where data are assimilated
close to the equator only.

 \begin{figure}[h!] 
 \centering
 \includegraphics[width=8cm]{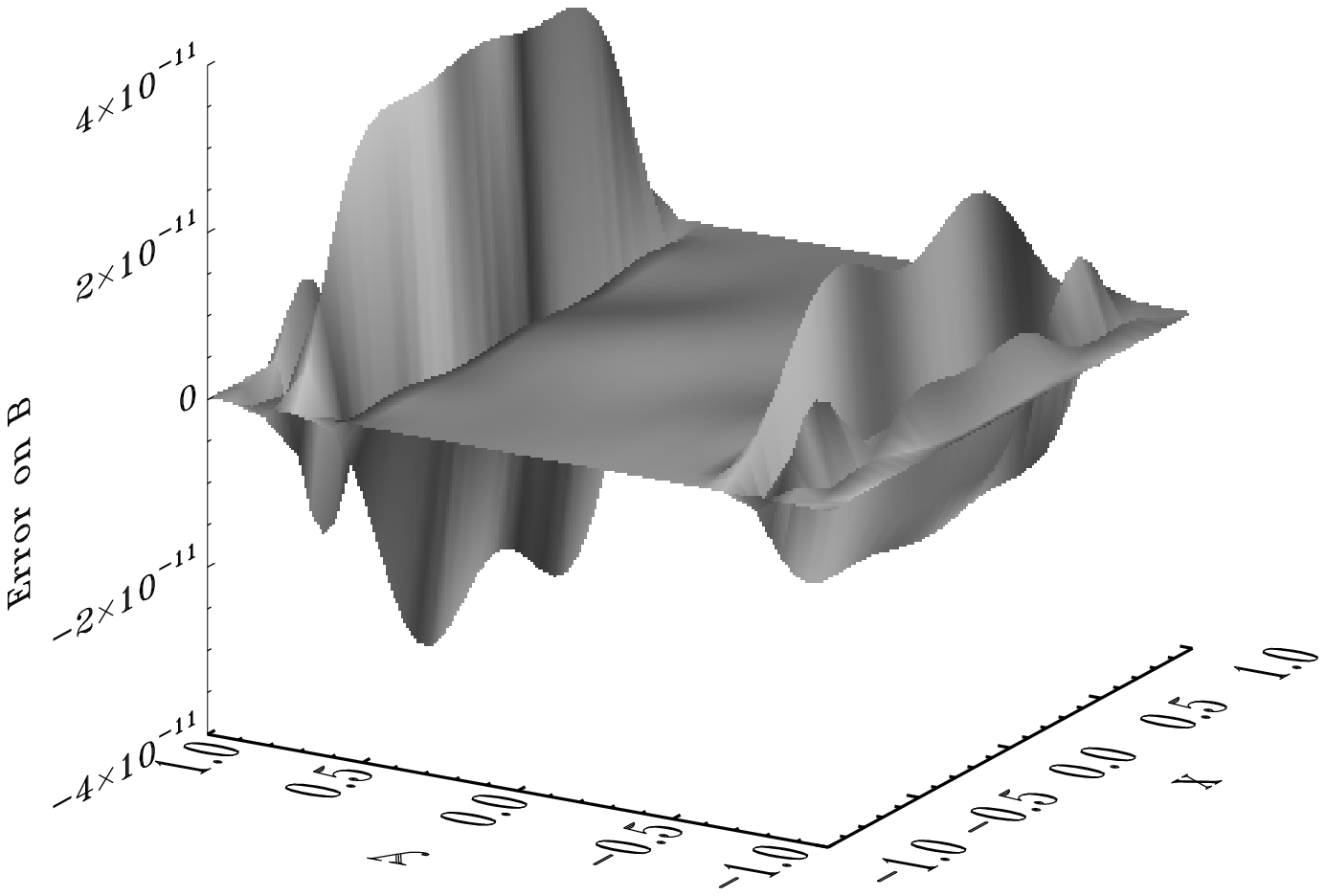}
 \includegraphics[width=8cm]{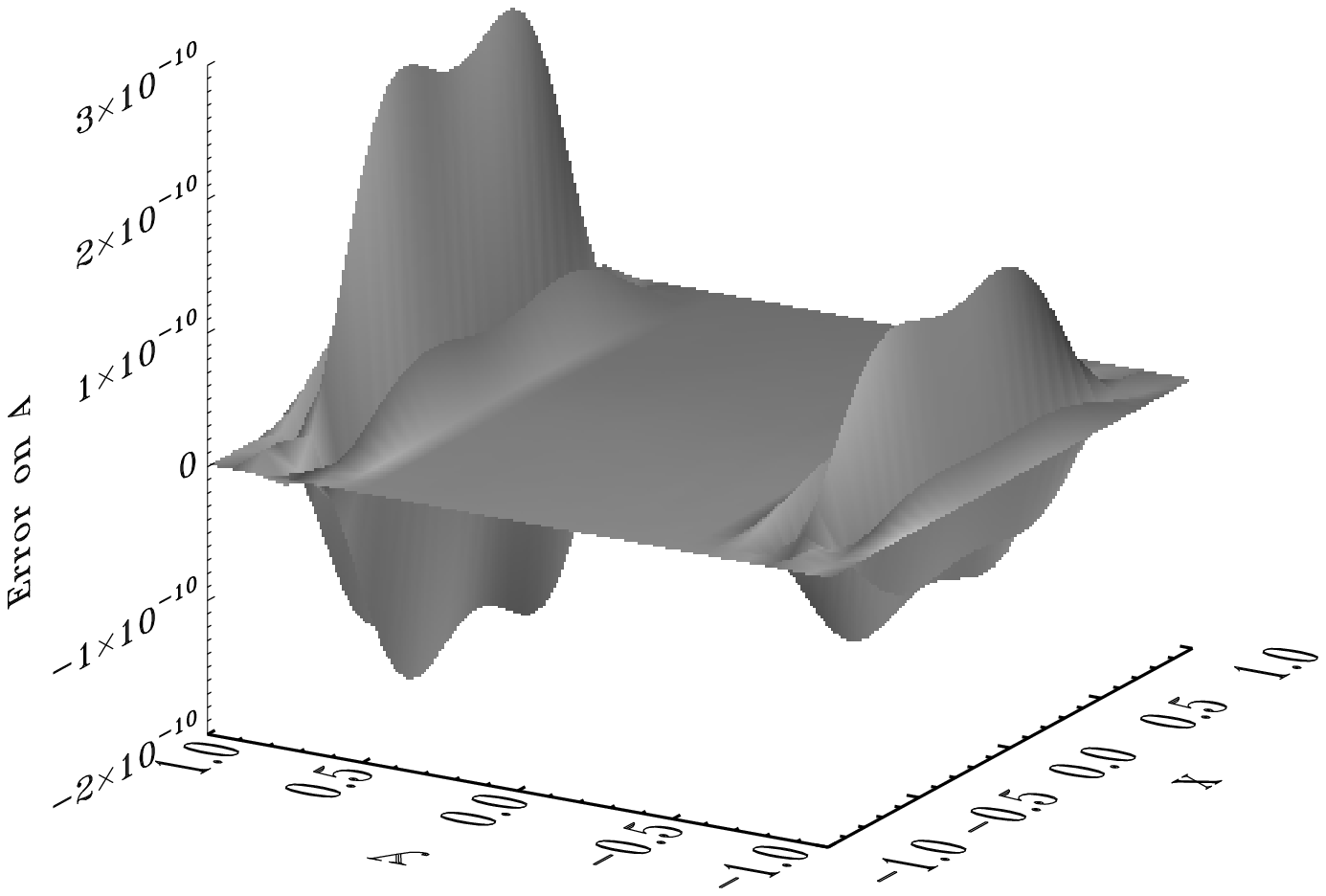}
 \caption{Same as Fig. \ref{fig_irreg} but for a case where data are
 assimilated in the equatorial region only between $-35^o$ and $35^o$.}
 \label{fig_error_act}
 \end{figure}

Figure \ref{fig_alpha_act1} shows the results of various experiments
where data have been assimilated in a more or less narrow band around
the equator. We present cases where observations
have been produced successively between $-18^o$
and $18^o$, $-26^o$ and $26^o$ and $-35^o$ and $35^o$. Figure
\ref{fig_alpha_act1} shows the difference of each reconstructed $\hat{\alpha}$ to the true
state. It is quite clear again that the recovery
of the correct $\hat{\alpha}$ is optimal at the locations where observations
were present. Indeed, the function is very smooth and close to the
true state at low latitudes for the first two runs. Close to the poles, the
fluctuations around the true $\hat{\alpha}$ can be quite significant, the error being there of the same
order as the function itself for the first run. However, when the area spanned by the observations increases,
the agreement with the true state improves and when observations
between about $-35^o$ and $35^o$ in latitude are used, the relative error made
on $\hat{\alpha}$ is as low as $10^{-7}$. This is an interesting property since the actual activity band in the Sun is approximately
located within those latitudes. We note in this particular case that
the errors are of the same order everywhere in the domain and that
the difference of knowledge/information between the region where data were available and the poles
is mostly absent. We conclude here that the whole function has been
recovered to a very good accuracy for this case where data were
assimilated only in the activity belt.

Once again, we can check the results on the magnetic fields produced
by the recovered $\hat{\alpha}$. Results are shown in Fig.
\ref{fig_error_act}. We chose to show the results for the assimilation
on the latitudinal band $-35^o$ to $35^o$ since the resulting $\hat{\alpha}$
function for this case was recovered to a very good and similar accuracy in the
whole domain. The largest errors both on the poloidal and toroidal fields at one instant are
again located mainly in the data-free regions. Nevertheless, we note that their amplitude remains very small,
even in the polar regions. Again, the errors on the poloidal field (for which we do not produce
observations) are about one order of magnitude larger than those on the toroidal field. It has to be noted here that the difference of knowledge/information
between the equator and the poles is visible, contrary to what
we found for the recovered $\hat{\alpha}$, stressing the not so direct
correspondence between the $\alpha$-effect and the magnetic field
evolution. The recovery within the equatorial band is
excellent, the error reaching values close to $10^{-12}$ for the
toroidal field and $10^{-11}$ for the poloidal field.


\subsection{Additional noise on the observed data}

In reality, the assimilated observations will always be contaminated by errors. Hence, it is
natural to study the behavior of our assimilation technique when
observations depart significantly from
what is directly produced by the numerical model. To do so, we produce
the same synthetic data by running the direct code once with the
choice of parameters quoted in sect.\ref{sect_model}. We
then add noise on the data by calculating

\begin{equation}
{B_z}_{noise}^{obs}=B_z^{obs}*(1+\sigma \, r)
\label{eq_noise}
\end{equation}

\noindent $r$ being a random number between $-1$ and $1$ and $\sigma$
measuring the departure from the synthetic data produced by the direct
code.

\begin{figure}[h!] 
 \centering
 \includegraphics[width=8cm]{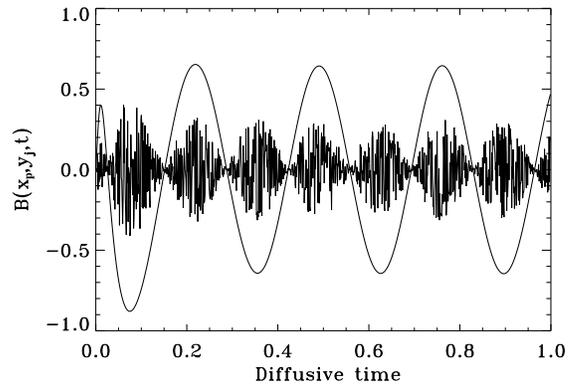}
 \caption{True state (smooth plain line) and error introduced in the data (magnified by a factor 50, fluctuating line) which will be
   used for the assimilation. This is a special case where the
   synthetic data have been perturbed by a noise with a standard
   deviation of $\sigma=10^{-2}$.}
 \label{fig_rand1}
 \end{figure}

As an illustration, we show in Fig. \ref{fig_rand1} the time
evolution of the ``true'' toroidal field at a specific point in space. We superimpose the error
on ${B_z}_{noise}^{obs}$ and the true state for $\sigma=10^{-2}$, magnified by a factor 50.
As a direct consequence of Eq. \ref{eq_noise}, the noise is
proportional to the value of $B_z$ and thus the errors are higher at periods of maximal activity.

\begin{figure} 
 \centering
 \includegraphics[width=8.2cm]{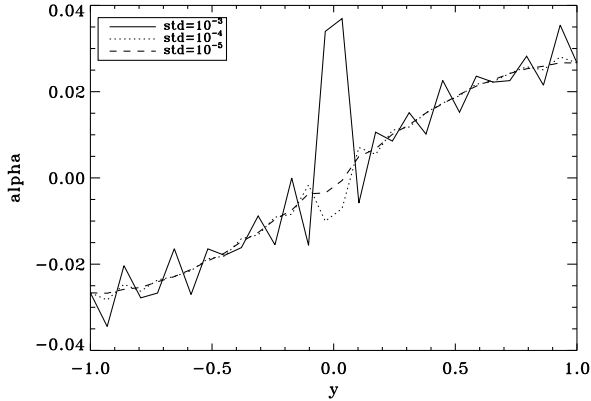}
 \caption{Reconstructed $\alpha$ after assimilation of data perturbed
   with random noises with various standard deviations. 100
   observations were used (10 in time and 10 along the y-direction).}
 \label{fig_rand2}
 \end{figure}

The results of the assimilation procedure are shown in Fig.
\ref{fig_rand2}. The number of observations used here was 100 (10 in
time multiplied by 10 in the y-direction). With the unperturbed
synthetic observations, the assimilation of those particular
observations gave us an $L_2$-error on $\alpha$ of about $6\times 10^{-8}$
and between $10^{-11}$ and $10^{-12}$ for the magnetic fields (see
figures \ref{fig_l2} and \ref{fig_error}). We will thus be able to
directly compare the results of the minimization after assimilation of
perturbed and unperturbed data. Four different experiments were
investigated, three of which are represented in Fig.
\ref{fig_rand2}. The only difference between those various experiments is the coefficient of the observation error $\sigma$.

From the figure, it is clear that the minimization of the cost
function gives $\hat{\alpha}$ profiles which agree less and less with the
true state when the noise on the assimilated data is increased. More
precisely, when $\sigma=10^{-5}$, the $\hat{\alpha}$ function is almost
perfectly recovered, except from a small region around the equator
in which the cost function is less sensitive to the values of
$\hat{\alpha}$. When $\sigma=10^{-4}$, the result of the minimization
procedure gives an $\hat{\alpha}$ which is already much less satisfactory, the 
$L_2$-error to the true state being of the order of $10^{-1}$ (compared
to $6\times 10^{-8}$ for the unperturbed case). When $\sigma$ is
further increased, the recovery of the $\hat{\alpha}$ profile is poor, the
error being of about $50\%$ in this case. The final errors on the
toroidal and poloidal magnetic fields are of the same order as the
errors introduced on the assimilated data, which shows that the minimization is fundamentally successful. Nevertheless, even if the true
state and the final fields depart of the same amount from the
perturbed observations, the errors between them are still significant. For $\sigma=10^{-5}$, the minimum $L_2$-error reached on $B_z$ is of
the order of $4\times10^{-5}$, about 6 orders of magnitude higher than
the typical errors in similar situations using unperturbed data.


\section{Conclusion}

We have presented the first attempt to apply variational data
assimilation techniques to the solar dynamo. A very simplified formulation
was used, namely a linear deterministic $\alpha\Omega$ dynamo model in Cartesian geometry, which
should not be taken as an accurate representation of the magnetic
field regeneration and evolution in the Sun. Nevertheless, we showed
that with this simple model, variational data assimilation gives us a
way to constrain various input parameters such as the profile of the
$\alpha$-effect, through the minimization of the errors to very few
observations (140 observations at most were used, out of 30000 points
in the (y,t) plane). With regularly-spaced
observations, the variational technique enabled us to recover the profile
of the $\alpha$-effect at the accuracy of about $10^{-8}$, starting
from an initial guess with an error of $10^{-2}$. This recovered
$\alpha$ then produced magnetic fields in extremely good agreement
(accuracy of around $10^{-12}$) with the true state.

Moreover, we showed that a partial knowledge of the toroidal field
could give us useful information on the $\alpha$-effect in the
  whole domain. Indeed, we showed that assimilating data
in the latitudinal belt of activity (between $-35^o$ and $35^o$) is
enough to reconstruct $\alpha$ at all latitudes with a final $L_2$-error of $10^{-7}$.
We also showed that adding noise on the observations strongly
perturbed the results of the minimization procedure, even if the global shape of
the $\alpha$-effect was mainly recovered in all cases (and especially the antisymmetry about the equator). Finally we showed that the reconstruction of the
$\alpha$-effect in our toy model is difficult near the equator if the observed (generated) data
assimilated in the procedure are insensitive to variations in that region, as it was the case here with $B_z$ being zero. 
However, if we had considered an $\alpha^2\Omega$ dynamo model (with the $\alpha$-effect present in the production of $B_z$), that comment might have not been true. This will be checked in future investigations.
Other quantities such as the differential rotation or observed variables such as the poloidal field could help better reconstructing information 
in these specific locations. It may then be worth trying several combinations of quantities and variables in our attempt to better determine the internal dynamics of the Sun.

The proof of concept presented in this work is very promising for the future developments of solar magnetic activity forecast. Indeed, we showed that if a
physical model is assumed to be sufficiently close to reality, the knowledge of a very small piece of information could
provide us with the reconstruction of a very important physical
process for which direct measurements are not available. More
precisely, in the case of the Sun, if we assume that the meridional
flow (large-scale flow in the meridional plane) plays a significant role in the evolution of the large-scale
magnetic field \cite{Dikpati06, Jouve07, Nandy11} and hence in the dynamo loop, data assimilation could
be very useful. Indeed, the meridional circulation is very difficult
to measure accurately, especially at depths higher than a few tens of
Mm (see review of Miesch 2005 and
recent observations of Hathaway \& Rightmire 2010). However, the magnetic field strength and configuration now start
to be detected with great accuracy through new satellites as Hinode
and SDO that provide vector magnetograms of the full solar disk. Data
assimilation is then a way to link the direct measurements of, say, the
radial field in active regions and a physical model in which the
meridional flow takes part in the dynamo loop. It is the case for
instance of flux-transport dynamo models which are sometimes used to
model the whole evolution of large-scale magnetic fields in the
Sun. Not only would some subtle physical processes (i.e. difficult to
detect directly) be reconstructed through the assimilation of accurate
observations of more accessible variables, but we could then use the
physical models to predict the behavior of the next solar cycle, with
a different technique from what was used up to now. 

We said in the
introduction that a reliable technique to predict future solar
magnetic phenomena still does not exist, we propose here a way to
progress in this direction, inspired by what has been used for a long
time already in the Earth weather community.  
Of course, better physical models and better understanding of
the physical processes interacting in a star need to be developed
before we can safely apply data assimilation techniques to give
tentative predictions of the solar activity. In particular, the goal
would be to assimilate observations of excellent quality (which are
already available) in 3D MHD global solar dynamo models producing
realistic magnetic cycles (which are not yet available). In the
meantime, we try to progress step by step towards this goal and we
think this work constitutes one of these steps, proving the
possibility to apply modern data assimilation techniques in solar physics.
A next step could be to use a nonlinear dynamo model that is sensitive to the initial conditions and
which uses polar coordinates rather than Cartesian ones. Finally, we could also introduce a so-called background term in the cost function, which limits the departure from an \emph{a priori} estimate of the state vector \cite[see][for further details]{Fournier10}. This allows to introduce data that is not contained in the observations such as information on the smoothness of the physical parameters (like the function $\alpha$ for example).
We intend to do so in the near future.

%
\begin{appendix}
   
\section{The adjoint induction equation}

In this section, we present in details the different steps leading to
the determination of the continuous adjoint mean-field induction equation. This
is only of particular use for the development of the adjoint model but
we find it useful to gain some insight on the link between adjoint
operators and the calculations shown in this work.

The velocity field ${\bf v}$, magnetic diffusivity  $\eta$ and the
$\alpha$-effect are given functions of space and time. We show here
how to compute the adjoint of each operator appearing in the
equation. We first define the adjoint operator in the following manner:\\


 $\Psi^{\star}$ is the adjoint of $\Psi$ operating on the Euclidian
 space $E$ if and only if

\begin{equation}
\forall ({\bf u_1},{\bf u_2}) \in E^2, \,\,\,\, \Psi ({\bf u_1})\cdot {\bf u_2} = {\bf u_1} \cdot \Psi^{\star} ({\bf u_2})
\label{eq_adjoint}
\end{equation}

\noindent where  $\cdot$ is the scalar product on $E$. As a
consequence, in order to determine the adjoint of an operator, we need to find the operator such that condition
\ref{eq_adjoint} is fulfilled.\\

Let ${\bf u_1}$ and ${\bf u_2}$ be elements of the Euclidian space $E$.\\

{\bf 1.} We first try to get the adjoint of the operator ${\bf u}
\rightarrow  {\bf v} \times {\bf u}$. Let  $ \Psi^{\star}$ be such
that $({\bf v} \times {\bf u_1}) \cdot {\bf u_2}={\bf u_1} \cdot
\Psi^{\star} ({\bf u_2})$. Then by manipulation of vector identities,
we get:

\begin{equation}
({\bf v}\times {\bf u_1}) \cdot {\bf u_2}=-{\bf u_1} \cdot ({\bf v}\times {\bf u_2})
\end{equation}

The adjoint of ${\bf u} \rightarrow  {\bf v} \times {\bf u}$ is thus ${\bf u} \rightarrow  -{\bf v} \times {\bf u}$. \\

{\bf 2.} Let us now look for the adjoint of ${\bf u} \rightarrow
\nabla \times {\bf u}$. Let $ \Psi^{\star}$ be such that $(\nabla \times {\bf u_1}) \cdot {\bf u_2}={\bf u_1}\cdot \Psi^{\star} ({\bf u_2})$.

\begin{equation}
(\nabla\times {\bf u_1}) \cdot {\bf u_2}={\bf u_1} \cdot (\nabla\times {\bf u_2})+ \nabla \cdot ({\bf u_1}\times {\bf u_2})
\end{equation}

The adjoint of ${\bf u} \rightarrow \nabla \times {\bf u}$ is thus
${\bf u} \rightarrow  \nabla \times {\bf u}$, the term $\nabla \cdot
({\bf u_1}\times {\bf u_2})$ representing a boundary term which will
be used in the adjoint integration to test the sensitivity of the cost
function to the boundary conditions for example.\\

{\bf 3.} W e now determine the adjoint of ${\bf u} \rightarrow  \alpha
{\bf u}$ Let $ \Psi^{\star}$ be such that $(\alpha {\bf u_1}) \cdot
{\bf u_2}={\bf u_1} \cdot \Psi^{\star} ({\bf u_2})$. It is
straightforward to see that

\begin{equation}
(\alpha {\bf u_1}) \cdot {\bf u_2}={\bf u_1} \cdot (\alpha {\bf u_2})
\end{equation}

The adjoint of ${\bf u} \rightarrow  \alpha {\bf u}$ is thus ${\bf u}
\rightarrow  \alpha {\bf u}$ (this operator is said to be self-adjoint). \\

{\bf 4.} Finally, we need to get the adjoint of ${\bf u} \rightarrow
\partial {\bf u}/\partial t$. Let $ \Psi^{\star}$ be such that
$\partial {\bf u_1}/\partial t \cdot {\bf u_2}={\bf u_1}\cdot
\Psi^{\star} ({\bf u_2})$. We have:

\begin{equation}
\frac{\partial {\bf u_1}}{\partial t} \cdot {\bf u_2}=-{\bf u_1}\cdot \frac{\partial {\bf u_2}}{\partial t} +\frac{\partial ({\bf u_1}\cdot {\bf u_2})}{\partial t}
\end{equation}

The adjoint of ${\bf u} \rightarrow  \partial {\bf u}/\partial t$ is
thus ${\bf u} \rightarrow  -\partial {\bf u}/\partial t$, the term
$\frac{\partial ({\bf u_1}\cdot {\bf u_2})}{\partial t}$ now
representing an initial conditions term which could be used the
adjoint integration to study the effect of the initial conditions on a
particular cost function.\\


We are thus able now to write the adjoint induction equation, using
the property that the adjoint of a composition of operators is the
compositions of the adjoint operators, taken in the reverse order.

Finally, we have: 

\begin{equation}
\frac{\partial {\bf B}}{\partial t}={\bf v} \times (\nabla \times {\bf B})-\alpha \nabla \times {\bf B}+\nabla\times(\eta\nabla{\bf \times B})
\label{eq_admf}
\end{equation}

\section{Variational approach}
\label{ap_B}

In this section we will follow and adapt the procedure described in \cite{Talagrand03}.
Let's consider the coupled induction equations \ref{eq_A_simple} and \ref{eq_B_simple} for the fields $A$ and $B_z$.
We search solutions of this set of equations over the rectangular domain $D=[x_1,x_2]\times[y_1,y_2]\times[t_1,t_2]$ in (x,y,t)-space. These equations are
first order with respect to $t$ and second order with respect to $x$ and $y$.

Consider now a field $B_z^{obs}(x,y,t)$ of observations over the domain $D$.
Since we assimilate data only on the toroidal field (as a proxy of the surface sunspots) our cost function $\jj$ is written:

\begin{equation}
 \jj(B) = \frac{1}{2}\iiint\limits_D (B_z - B_z^{obs})^2 \, dx dy dt
\end{equation}

its variation is thus:

\begin{equation}
 \dj=\intt (B_z - B_z^{obs}) \db \, dx dy dt
\end{equation}

We aim at expressing the variations of the cost function $\jj$ to
  variations of our well-defined input parameters which are:

\begin{itemize}

\item[-] {The values of $A$ and $B_z$ for all points in space at the
  initial time $t=t_1$}
\item[-] {The constant magnetic diffusivity $\eta$}
\item[-] {The function representing the $\alpha$-effect $\alpha(x,y)$}
\item[-] {The azimuthal velocity function $v(x,y)$}

\end{itemize}

Let's derive the {\it tangent linear equation} obtain by differentiating equations \ref{eq_A_simple} and \ref{eq_B_simple} with respect to $A$, $B_z$, the parameter $\eta$ and the functions $v$ and $\alpha$
and name respectively their variations $\da$, $\db$, $\de$, $\dv$, $\dh$. The equations read:
\begin{equation}
\frac{\partial \da}{\partial t}-\dh B_z - \alpha \db - \eta (\frac{\partial^2 \da}{\partial x^2}+ \frac{\partial^2 \da}{\partial y^2}) - \de (\frac{\partial^2 A}{\partial x^2}+ \frac{\partial^2 A}{\partial y^2} )= 0
\label{eq_A_tl}
\end{equation}

\begin{eqnarray}
\frac{\partial \db}{\partial t}&-&\frac{\partial \dv}{\partial x}\frac{\partial A}{\partial y}+\frac{\partial \dv}{\partial y}\frac{\partial A}{\partial x}-\frac{\partial v}{\partial x}\frac{\partial \da}{\partial y}+\frac{\partial v}{\partial y}\frac{\partial \da}{\partial x} \nonumber \\
&-&\eta (\frac{\partial^2 \db}{\partial x^2}+ \frac{\partial^2 \db}{\partial y^2})-\de (\frac{\partial^2 B_z}{\partial x^2}+ \frac{\partial^2 B_z}{\partial y^2}) = 0
\label{eq_B_tl}
\end{eqnarray}


Using Lagrange multipliers $\lambda(x,y,t)$ and $\gamma(x,y,t)$ respectively for equations \ref{eq_A_tl} and \ref{eq_B_tl}, we get (introducing a negative sign for
simplicity):

\begin{flalign}
\dj=\intt \left( (B_z - B_z^{obs}) \db - \lambda \left [ \frac{\partial \da}{\partial t}-\dh B_z - \alpha \db - \eta (\frac{\partial^2 \da}{\partial x^2}+ \frac{\partial^2 \da}{\partial y^2}) \nonumber \right. \right. \\ \left.
- \de (\frac{\partial^2 A}{\partial x^2}+ \frac{\partial^2 A}{\partial y^2} ) \right] -\gamma \left[ \frac{\partial \db}{\partial t}-\frac{\partial \dv}{\partial x}\frac{\partial A}{\partial y}+\frac{\partial \dv}{\partial y}\frac{\partial A}{\partial x}-\frac{\partial v}{\partial x}\frac{\partial \da}{\partial y}+\frac{\partial v}{\partial y}\frac{\partial \da}{\partial x} \right. \\ \left. \left. 
-\eta (\frac{\partial^2 \db}{\partial x^2}+ \frac{\partial^2 \db}{\partial y^2})-\de (\frac{\partial^2 B_z}{\partial x^2}+ \frac{\partial^2 B_z}{\partial y^2})\right ] \right) \,dx dy dt  \nonumber
\end{flalign}

We now wish to remove all the differentiation operating on $A$, $B_z$, $\eta$, $v$ and $\alpha$. To do so we use as many integration by parts as necessary.
for the sake of clarity we demonstrate the procedure for a few typical terms:

\begin{eqnarray}
-\intt \lambda \frac{\partial \da}{\partial t} \, dx dy dt = -\intxy \lambda \da \vtt + \intt \frac{\partial \lambda}{\partial t} \da \, dx dy dt 
\end{eqnarray} 

Diffusion terms require a double integration by parts:
\begin{eqnarray}
\intt \lambda \eta \frac{\partial^2 \da}{\partial x^2} \, dx dy dt &=& \intyt \lambda \eta \frac{\partial \da}{\partial x} \vtx - \intt \frac{\partial (\lambda \eta)}{\partial x} \frac{\partial \da}{\partial x} \, dx dy dt \nonumber \\
&=&\intyt \lambda \eta \frac{\partial \da}{\partial x} \vtx- \intyt \frac{\partial (\lambda \eta)}{\partial x} \da \vtx \nonumber \\
&+& \intt \frac{\partial^2 (\lambda \eta)}{\partial x^2} \da \,dx dy dt 
\end{eqnarray} 
We note that the double integration modifies the sign of the diffusion term relative to the time derivative, this is expected since by going backward in time
we should anti-diffuse.

\begin{eqnarray}
-\intt \gamma \frac{\partial v}{\partial y} \frac{\partial \da}{\partial x} \, dx dy dt = -\intyt \gamma \frac{\partial v}{\partial y} \da \vtx + \intt \frac{\partial}{\partial x} (\gamma \frac{\partial v}{\partial y}) \da \, dx dy dt 
\end{eqnarray} 

Terms involving variations of the ingredients are treated likewise:
\begin{eqnarray}
-\intt \gamma \frac{\partial A}{\partial x} \frac{\partial \dv}{\partial y} \, dx dy dt = -\intxt \gamma \frac{\partial A}{\partial x} \dv \vty + \intt \frac{\partial}{\partial y} (\gamma \frac{\partial A}{\partial x}) \dv \,dx dy dt 
\end{eqnarray} 

Applying systematically this method to all the terms that possess differentiation of the variations and grouping the terms by variations, we get for $\dj$ the following equation:

\begin{flalign}
\label{djj}
\dj=\intt \left( \left[ \frac{\partial \lambda}{\partial t} + \eta (\frac{\partial^2 \lambda}{\partial x^2}+ \frac{\partial^2 \lambda}{\partial y^2})+\frac{\partial}{\partial x}(\gamma \frac{\partial v}{\partial y})-\frac{\partial}{\partial y}(\gamma \frac{\partial v}{\partial x}) \right]\da  \right. \nonumber \\ \left. 
+ \left[ \frac{\partial \gamma}{\partial t} + \lambda \alpha + \eta (\frac{\partial^2 \gamma}{\partial x^2}+ \frac{\partial^2 \gamma}{\partial y^2}) + (B_z - B_z^{obs}) \right ]\db  + \dh \lambda B_z \right. \nonumber \\ \left.
+ \de \left[\lambda(\frac{\partial^2 A}{\partial x^2}+ \frac{\partial^2 A}{\partial y^2} ) + \gamma (\frac{\partial^2 B_z}{\partial x^2}+ \frac{\partial^2 B_z}{\partial y^2})\right] \right. \nonumber \\ \left.
-\dv\left[\frac{\partial}{\partial x}(\gamma \frac{\partial A}{\partial y})-\frac{\partial}{\partial y}(\gamma \frac{\partial A}{\partial x})\right] \right) \, dx dy dt \nonumber \\
- \intxy \lambda \da \vtt \nonumber \\
+ \intyt \left(\eta\left[\lambda \frac{\partial \da}{\partial x}- \frac{\partial \lambda}{\partial x}\da \right] - \gamma \frac{\partial v}{\partial y}\da + \gamma \dv \frac{\partial A}{\partial y}\right)  \vtx  \nonumber \\ 
+ \intxt \left(\eta\left[\lambda \frac{\partial \da}{\partial y}- \frac{\partial \lambda}{\partial y}\da \right] + \gamma \frac{\partial v}{\partial x}\da - \gamma \dv \frac{\partial A}{\partial x}\right)  \vty  \nonumber \\ 
- \intxy \gamma \db \vtt  + \intyt \eta\left[\gamma \frac{\partial \db}{\partial x}- \frac{\partial \gamma}{\partial x}\db \right] \vtx \nonumber \\
+ \intxt \eta\left[\gamma \frac{\partial \db}{\partial y}- \frac{\partial \gamma}{\partial y}\db \right] \vty 
\end{flalign}


This expression is valid for any $\lambda(x,y,t)$ and $\gamma(x,y,t)$. 
The space-time integral can be cancelled by imposing that $\lambda$ and $\gamma$ verify the following partial differential equations: 

\begin{eqnarray}
\frac{\partial \lambda}{\partial t} &+& \eta (\frac{\partial^2 \lambda}{\partial x^2}+ \frac{\partial^2 \lambda}{\partial y^2})+\frac{\partial}{\partial x}(\gamma \frac{\partial v}{\partial y})-\frac{\partial}{\partial y}(\gamma \frac{\partial v}{\partial x}) = 0 
\label{eq_gl1}
\end{eqnarray}
\begin{eqnarray}
\frac{\partial \gamma}{\partial t} &+& \lambda \alpha + \eta
(\frac{\partial^2 \gamma}{\partial x^2}+ \frac{\partial^2
  \gamma}{\partial y^2}) + (B_z - B_z^{obs}) = 0
\label{eq_gl2}
\end{eqnarray}

We also impose that $\lambda(x,y,t_2)$ and $\gamma(x,y,t_2)$ equal zero for any x or y. Further since $A$ and $B_z$ are constrained to be equal to zero at the boundary, their variation are zero, and all the terms involving either $\da$ and $\db$ in the surface integrals above vanish. 
For the terms involving derivatives of $\da$ and $\db$ we impose that
$\lambda$ and $\gamma$ equal zero on boundaries $x=x_1,x_2$ and
$y=y_1,y_2$ for any t.

We note that equations \ref{eq_gl1} and \ref{eq_gl2} with the conditions on $\lambda$
and $\gamma$ just stated above unambiguously define the functions in
the whole domain. Equations \ref{eq_gl1} and \ref{eq_gl2} are first-order in time and
second-order in space, and define a well-posed problem for backward
integration with respect to $t$, because of the positive sign of the
diffusion terms. The specification of $\lambda$ and $\gamma$ at the
final time $t_2$ and along the spatial boundaries $x_1, x_2, y_1$ and
$y_2$ therefore unambiguously define the functions $\lambda(x,y,t)$
and $\gamma(x,y,t)$. 

Taking into account these various conditions, equation \ref{djj} reduces to:

\begin{multline}
\dj=\intxy \lambda(x,y,t_1) \da(x,y,t_1) dxdy +\intxy \gamma(x,y,t_1) \db(x,y,t_1) dxdy +\intxy \left[ \dh \int_{t_1}^{t_2} \lambda B_z dt \right] dx dy \\
+ \de \intt \left[\lambda(\frac{\partial^2 A}{\partial x^2}+ \frac{\partial^2 A}{\partial y^2} ) + \gamma (\frac{\partial^2 B_z}{\partial x^2}+ \frac{\partial^2 B_z}{\partial y^2})\right]\, dx dy dt  
- \intt \dv\left[\frac{\partial}{\partial x}(\gamma \frac{\partial A}{\partial y})-\frac{\partial}{\partial y}(\gamma \frac{\partial A}{\partial x})\right] \, dx dy dt
\end{multline}


The above equation reveals that the partial derivatives of the objective function $\jj$ with respect to $\eta$, $\alpha(x,y)$, $v(x,y)$, $A(x,y,t_1)$ and $B_z(x,y,t_1)$ are equal to:

\begin{eqnarray}
\frac{\partial \jj}{\partial A}(x,y,t_1)&=&\lambda(x,y,t_1) \, \mbox{ and }\,  \frac{\partial \jj}{\partial B_z}(x,y,t_1)=\gamma(x,y,t_1)  \\
\frac{\partial \jj}{\partial \eta}&=&\intt \left[\lambda(\frac{\partial^2 A}{\partial x^2}+ \frac{\partial^2 A}{\partial y^2} ) + \gamma (\frac{\partial^2 B_z}{\partial x^2}+ \frac{\partial^2 B_z}{\partial y^2})\right] \, dx dy dt  \\
\frac{\partial \jj}{\partial \alpha}&=&\int_{t_1}^{t_2}\lambda B_z \, dt \, \,\,\,\, , \forall{(x,y)} \label{eq_dj} \\
\frac{\partial \jj}{\partial v}&=&\int_{t_1}^{t_2} \left[ \frac{\partial}{\partial y}(\gamma \frac{\partial A}{\partial x})-\frac{\partial}{\partial x}(\gamma \frac{\partial A}{\partial y})\right] \, dt \,\,\,\,\, , \forall{(x,y)} 
\end{eqnarray}
We here solve a simplified problem by  considering that $A$ and $B_z$ at $t=t_1$ are known \cite[see][for discussions about sensitivity to initial conditions]{Talagrand87}.

\end{appendix}

%
 \begin{acknowledgements}
We thank A. Fournier, A. Vincent, E. Canet, D. Jault, S. Kosovichev, M. DeRosa, M. Dikpati, P. Gilman and I. Kitiashvili for fruitful discussions 
and for sharing their own experience in using data assimilation techniques for geophysical and solar physic problems. A.S. Brun and L. Jouve acknowledge 
financial support by the ERC starting grant 207430 STARS2 and by the CNRS/INSU Programme National Soleil-Terre. All authors are thankful to ISSI for hosting
our international group on data assimilation.
 \end{acknowledgements}

%
%

\begin{thebibliography}{}

\bibitem[{Archontis et al.} {2005}]{Archontis05} Archontis, V., 
Moreno-Insertis, F., Galsgaard, K., \& Hood, A.~W.\ 2005, \apj, 635,
1299 
\bibitem[{Beer et al.} {1998}]{Beer98} Beer, J., Tobias, S., 
\& Weiss, N.\ 1998, Solar Physics, 181, 237 
\bibitem[{B{\'e}langer et al.} {2005}]{Belanger05} B{\'e}langer, E., 
Charbonneau, P., \& Vincent, A.\ 2005, Journal of the Royal Astronomical Society of Canada, 99, 133 
\bibitem[{Bocquet} {2011}]{Bocquet11} Bocquet, M. \ 2011, Notes de cours du M2 OACOS, de l'ENSTA et de l'Ecole des Ponts ParisTech
\bibitem[{Browning et al.} {2006}]{Browning06} Browning, M.~K., 
Miesch, M.~S., Brun, A.~S., \& Toomre, J.\ 2006, \apjl, 648, L157 
\bibitem[{Brun et al.} {2004}]{Brun04}Brun, A.S., Miesch, S.M. \&
  Toomre, J. \ 2004, ApJ, 614, 1073
\bibitem[{Brun et 
al.} {2011}]{Brun11} Brun, A.~S., Miesch, S.M. \&
  Toomre, J. \ 2011, \apj, submitted
\bibitem[{Cameron 
\& Sch{\"u}ssler} {2007}]{Cameron07} Cameron, R., \& Sch{\"u}ssler, M.\ 2007, \apj, 659, 801 
\bibitem[{Cattaneo \& Hughes} {2001}]{Cattaneo01}Cattaneo, F., Hughes, D.W., 2001, A\&G, 42, 18 
\bibitem[{Charbonneau} {2005}]{Charbonneau05}Charbonneau, P., 2005, Living Rev. Solar
  Phys., 2
\bibitem[{Choudhuri et al.} {2007}]{Choudhuri07} Choudhuri, A.~R., 
Chatterjee, P., \& Jiang, J.\ 2007, Physical Review Letters, 98, 131103 
\bibitem[{Cline} {2003}]{Cline03} Cline, K.~S.\ 2003, Ph.D.~Thesis
\bibitem[{Daley} {1991}]{Daley91} Daley, R.\ 1991, Science, 254, 
1531 
\bibitem[{deJager \& Duhau} {2009}]{deJager09} de Jager \& Duhau,
  S. 2009, JAST, 71, 239 
\bibitem[{Dikpati et al.} {2004}]{Dikpati04} Dikpati, M., de Toma, 
G., Gilman, P.~A., Arge, C.~N., \& White, O.~R.\ 2004, \apj, 601, 1136 
\bibitem[{Dikpati \& Gilman} {2006}]{Dikpati06} Dikpati, M., \&
  Gilman, P.~A.\ 2006, \apj, 649, 498 
\bibitem[{Dikpati et al.} {2006}]{Dikpati062} Dikpati, M., de Toma, 
G., \& Gilman, P.~A.\ 2006, \grl, 33, 5102 
\bibitem[{Duhau} {2003}]{Duhau03} Duhau, S.\ 2003, \solphys, 213, 
203 
\bibitem[{Fan et al.} {2003}]{Fan03}Fan, Y., Abbett, W.P. \& Fisher,
  G. H. 2003, \apj, 582, 1206
\bibitem[{Fournier et al.}{, 2007}]{Fournier07}Fournier, A., Eymin, C. \&
  Alboussi\`ere, T.\ 2007, Nonlin. Processes Geophys., 14, 163
\bibitem[{Fournier et al.} {2010}]{Fournier10} Fournier, A., et al.\ 
2010, \ssr, 155, 24
\bibitem[{Giering \& Kaminski} {1998}]{Giering98} Giering, R. \& Kaminski, T.\ 1998, ACM Transactions on Mathematical Software, 24, 437
\bibitem[{Hathaway} {2010}]{Hatha10} Hathaway, D.~H.\ 2010, Living review in solar physics,  7, 1, http://solarphysics.livingreviews.org/Articles/lrsp-2010-1/
\bibitem[{Hathaway et al.} {1999}]{Hathaway99} Hathaway, D.~H., 
Wilson, R.~M., \& Reichmann, E.~J.\ 1999, \jgr, 104, 22375 
\bibitem[{Hathaway 
\& Wilson} {2004}]{Hathaway04} Hathaway, D.~H., \& Wilson, R.~M.\ 2004, 
\solphys, 224, 5 
\bibitem[{Hathaway 
\& Rightmire} {2010}]{Hathaway10} Hathaway, D.~H., \& Rightmire, L.\ 2010, Science, 327, 1350 
\bibitem[{Jouve 
\& Brun} {2007}]{Jouve07} Jouve, L., \& Brun, A.~S.\ 2007, \aap, 474, 239 
\bibitem[{Jouve 
\& Brun} {2009}]{Jouve09} Jouve, L., \& Brun, A.~S.\ 2009,
Astrophysical Journal, 701, 1300 
\bibitem[{Kalnay} 2003]{Kalnay03}Kalnay, E. \ 2003, Atmospheric
  Modeling, Data Assimilation and Predictability. Cambridge Press,
  341pp.
\bibitem[{Kitiashvili 
\& Kosovichev} {2008}]{Kitiashvili08} Kitiashvili, I., \& Kosovichev, A.~G.\ 2008, \apjl, 688, L49
\bibitem[{Komm et al.} {2007}]{Komm07} Komm, R., Howe, R., Hill, 
F., Miesch, M., Haber, D., \& Hindman, B.\ 2007, \apj, 667, 571 
 \bibitem[{Komm et al.} {2008}]{Komm08} Komm, R., Hill, F., 
\& Howe, R.\ 2008, Journal of Physics Conference Series, 118, 012035 
\bibitem[{Krause \&  Raedler}{, 1980}]{Krause80}Krause, F., \&
  Raedler, K.-H.\ 1980, Oxford, Pergamon Press, Ltd., 1980.~271 p.
  \bibitem[{Le Dimet 
\& Talagrand} {1986}]{LeDimet86} Le Dimet, F.-X., \& Talagrand, O.\ 1986, Tellus Series A, 38, 97
\bibitem[Lorenc {1981}]{Lorenc81} Lorenc, A.~C.\ 1981, Monthly 
Weather Review, 109, 701 
\bibitem[{Magara 
\& Longcope} {2003}]{Magara03} Magara, T., \& Longcope, D.~W.\ 2003, \apj, 586, 630 
\bibitem[{Miesch et al.} {2000}]{Miesch00} Miesch, M.~S., Elliott, 
J.~R., Toomre, J., Clune, T.~L., Glatzmaier, G.~A., 
\& Gilman, P.~A.\ 2000, \apj, 532, 593 
\bibitem[{Miesch} {2005}]{Miesch05}Miesch, M.~S.\ 2005, Living Reviews
  in Solar Physics, 2, 1
\bibitem[{Moffatt} {1978}]{Moffatt78} Moffatt, H.~K.\ 1978, 
Cambridge, England, Cambridge University Press, 1978.~353 p.
\bibitem[{Nandy et al.} {2011}]{Nandy11} Nandy, D., 
Mu{\~n}oz-Jaramillo, A., \& Martens, P.~C.~H.\ 2011, \nat, 471, 80
\bibitem[{Ossendrijver} {2003}]{Ossendrijver03}Ossendrijver,M., 2003,
  Astronomy and Astrophysics Review, 324, 64
\bibitem[{Pevtsov 
\& Canfield} {2001}]{Pevtsov01} Pevtsov, A.~A., \& Canfield, R.~C.\ 2001, \jgr, 106, 25191 
\bibitem[{Polak}{, 1971}]{Polak71}Polak,E.\ 1971, Computational Methods
  in Optimization, New York Academic Press, pp.56ff.
\bibitem[{Rempel \& Dikpati} {2009}]{Rempel09} Rempel, M \& Dikpati, M., ASPS, 416, 551  in Solar-stellar dynamos as revealed by helio- and asteroseismology, Gong2008/Soho21
\bibitem[{Roth} {2009}]{Roth09} Roth 2009, ASPS, 416, 501, in
  Solar-stellar dynamos as revealed by helio- and asteroseismology,
  Gong2008/Soho21
  \bibitem[{Schrijver \& DeRosa} {2003}]{Schrijver03} Schrijver, C. \& DeRosa, M. \ 2003, Solar Physics, 212, 165 
\bibitem[{Schwenn} {2006}]{Schwenn06} Schwenn, R.\ 2006, Living 
Reviews in Solar Physics, 3, 2 
\bibitem[{Steenbeck \& Raedler}{, 1966}]{Steenbeck66}Steenbeck, M., 
  Krause, F., Raedler, K.-H.\ 1966, Zeitschrift Naturforschung Teil A,
  21, 369 
\bibitem[{Stix} {2002}]{Stix02}Stix, M., 2002, {\it The Sun: an
    introduction}, Springer
\bibitem[{Svalgaard et al.} {2005}]{Svalgaard05} Svalgaard, L., 
Cliver, E.~W., \& Kamide, Y.\ 2005, \grl, 32, 1104 
\bibitem[{Talagrand 
\& Courtier} {1987}]{Talagrand87} Talagrand, O., \& Courtier, P.\ 1987, Quarterly Journal of the Royal Meteorological Society, 113, 1311 
\bibitem[{Talagrand} {1991}]{Talagrand91}Talagrand, O.\ 1991,
  Automatic Differentiation of Algorithms, Proceedings, A. Griewank
  and G.G. Corliss, editors, Society for Industrial and Applied
  Mathematics, Philadelphia
\bibitem[{Talagrand} {1997}]{Talagrand97}Talagrand, O. \ 1997,  Assimilation of observations, an introduction. J. Meteorol. SOC.Japan 75, 191-209.
\bibitem[{Talagrand} {2003}]{Talagrand03}Talagrand, O.\ 2003, Data
  Assimilation for the Earth System, Proceedings, Advanced Study
  Insitute, Acquafredda di Maratea, Italy, May-June 2002, Kluwer
  Academic Publishers, Dordrecht, The Netherlands, 37-53 
\bibitem[{Yoshimura} {1975}]{Yoshimura75}Yoshimura, H.\ 1975,
  Astrophysical Journal, 201, 740 
\bibitem[{Wang \& Sheeley} {1991}]{Wang91} Wang, Y.-M. \& Sheeley, N.R. 1991, ApJ, 375, 761

 \end{thebibliography}
%

%
%
%

\end{document}